\documentclass[journal,10pt]{IEEEtran}
\usepackage{lineno}
\usepackage{cite}
\usepackage{hyperref}
\usepackage{amsmath,amssymb,amsfonts}
\usepackage{amsthm}
\usepackage{mdwmath}
\usepackage{mdwtab}
\DeclareMathOperator*{\argmax}{argmax}

\usepackage{graphicx}
\usepackage{textcomp}
\usepackage{xcolor}
\usepackage{float}
\usepackage{subfigure}
\usepackage{mathrsfs}
\usepackage{mathtools}
\usepackage{multirow}
\usepackage{algpseudocode}
\usepackage[linesnumbered,ruled,vlined]{algorithm2e}
\usepackage[justification=centering]{caption}
\usepackage{setspace}
\usepackage{footmisc}
\usepackage{caption} 
\usepackage{dsfont}
\usepackage{bbm}
\usepackage{cases}
\usepackage{stfloats}

\theoremstyle{definition}
\newtheorem{remark}{Remark}
\newtheorem{corollary}{Corollary}

\newtheorem{theorem}{Theorem}

\newtheorem{lemma}{Lemma}

\makeatletter
\newcommand{\biggg}{\bBigg@{3}}
\newcommand{\Biggg}{\bBigg@{3.5}}
\makeatother
\usepackage{bm}
\begin{document}
\title{{Modeling and Analysis of Near-Field ISAC}}

\author{Boqun~Zhao,~\IEEEmembership{Graduate Student Member,~IEEE}, Chongjun~Ouyang,~\IEEEmembership{Member,~IEEE},
        Yuanwei~Liu,~\IEEEmembership{Fellow,~IEEE},
        Xingqi~Zhang,~\IEEEmembership{Member,~IEEE},
        and H.~Vincent~Poor,~\IEEEmembership{Life Fellow,~IEEE}
\thanks{This work was supported in part by the U.S. National Science Foundation under Grant CNS-2128448. An earlier version of this paper is to be presented in part at the IEEE International Conference on Communications, Denver, CO, USA, June 2024 \cite{ICC_2024}. \textit{(Corresponding authors: Yuanwei Liu, Xingqi Zhang.)}}
\thanks{B. Zhao and X. Zhang are with the Department of Electrical and Computer Engineering, University of Alberta, Edmonton, T6G 2H5, Canada (email:\{boqun1, xingqi.zhang\}@ualberta.ca).}
\thanks{C. Ouyang is the School of Electrical and Electronic Engineering, College of Engineering and Architecture, University College Dublin, Dublin 4, D04 V1W8, Ireland, and also with the Department of Electronic Engineering and Computer Science, Queen Mary University of London, E1 4NS London, U.K. (e-mail: chongjun.ouyang@ucd.ie).}
\thanks{Y. Liu is  with the School of Electronic Engineering and Computer Science, Queen Mary University of London, London, E1 4NS, U.K. (email: yuanwei.liu@qmul.ac.uk).}
\thanks{H. V. Poor is with the Department of Electrical and Computer Engineering, Princeton University, Princeton, NJ 08544, USA (email: poor@princeton.edu).}
}

\maketitle

\begin{abstract}
As the technical trends for the next-generation wireless network significantly extend the near-field region, a performance reevaluation of integrated sensing and communications (ISAC) with an appropriate channel model to account for the effects introduced by the near field becomes essential. In this paper, a near-field ISAC framework is proposed for both downlink and uplink scenarios based on an accurate channel model. A uniform planar array is equipped at a base station, where the impacts of the effective aperture and polarization of antennas are considered. For the downlink case, three distinct designs are studied: {a} communications-centric (C-C) design, {a} sensing-centric (S-C) design, and {a} Pareto optimal design. Regarding the uplink case, the C-C design, the S-C design and {a} time-sharing strategy are considered. Within each design, sensing rates (SRs) and communication rates (CRs) are derived. To gain further insights, high signal-to-noise ratio slopes and rate scaling laws concerning the number of antennas are examined. The attainable near-field SR-CR regions of ISAC and the baseline frequency-division S\&C are also characterized. Numerical results reveal that, as the number of antennas in the array grows, the SRs and CRs under our accurate model converge to finite values, while those under conventional far- and near-field models exhibit unbounded growth, highlighting the importance of precisely modeling the channels for near-field ISAC.
\end{abstract}


\section{Introduction}
The Integrated Sensing and Communications (ISAC) concept has generated substantial interest among researchers and the industry, owing to its promise in sixth-generation (6G) and {other emerging} wireless networks \cite{Zhang2021_JSTSP}. ISAC stands out due to its unique capability to efficiently share time, frequency, power, and hardware resources for both communication and sensing tasks simultaneously. This distinguishes it from the conventional approach known as Frequency-Division Sensing and Communications (FDSAC), where distinct frequency bands and infrastructure are needed for each function. ISAC is anticipated to deliver superior efficiency compared to FDSAC in terms of spectrum utilization, energy efficiency, and hardware demands \cite{ISACoverview_3,LiuAn}. Furthermore, ISAC can also be combined with other emerging techniques, such as reconfigurable surface \cite{star_1,star_2}, providing additional improvements in the performance of sensing and communications (S\&C) \cite{starISAC}. Therefore, ISAC holds significant research value.

In evaluating the effectiveness of ISAC, two essential performance metrics are commonly used from an information-theoretic perspective: sensing rate (SR) and communication rate (CR) \cite{LiuAn,Tang2019_TSP}. SR measures the system's ability to estimate environmental information through sensing processes, while CR quantifies the system's capacity for efficient data transmission during communication process. Analyzing these two metrics provides valuable insights into the overall performance and effectiveness of ISAC in seamlessly integrating S\&C functions.

In light of recent developments in wireless S\&C, there is a growing need to accommodate the demanding requirements of next-generation wireless networks. This entails the application of extremely large-scale antenna arrays and tremendously high frequencies \cite{NFC,Xidong_magazine}. With these technical trends, according to the merit for distinguishing between the near-field and far-field regions, i.e., Rayleigh distance $\frac{2D^2}{\lambda}$ with $D$ denoting the antenna aperture and $\lambda$ denoting the wavelength \cite{rayleighdis}, the region of near field will significantly expand, encompassing distances of several hundred meters. For example, consider a large-scale array with an aperture size of $D = 0.5$ meters operating at a frequency of 60 GHz, which results in a near-field region of $100$ meters. It is crucial to emphasize that electromagnetic (EM) waves exhibit distinct propagation characteristics in the near-field region as compared to the far field. In the far-field region, EM waves can be adequately approximated as planar waves. However, in the near field, a more precise modeling approach is required, involving spherical waves. Consequently, the conventional uniform planar wave (UPW) model employed in existing works analyzing the performance of ISAC within the far-field region, e.g., \cite{ISAC_performance1,ISAC_performance3,Ouyang2022_WCL,boqun_NOMAISAC}, is no longer valid in the near-field region. Therefore, it is imperative to reevaluate the performance of ISAC systems from a near-field perspective.

So far, there has been limited existing literature addressing near-field ISAC \cite{NFISAC_overview2,NFISAC_1,NFISAC_2,NFISAC_3,NFISAC_4}. In \cite{NFISAC_overview2}, the authors provided overviews of the effects of the near field on ISAC and explored the potential of near-field ISAC. The works of \cite{NFISAC_1,NFISAC_2,NFISAC_3,NFISAC_4} primarily focused on waveform or beamforming design aspects for the downlink near-field ISAC, where \cite{NFISAC_1,NFISAC_2} considered the uniform spherical wave (USW) and \cite{NFISAC_3,NFISAC_4} considered the non-uniform spherical wave (NUSW). However, the performance of near-field ISAC regarding both downlink and uplink case remains unexplored. 

Furthermore, despite the adoption of spherical wave models in the existing studies of near-field ISAC, both the USW model and the more precise NUSW model are still considered unreasonable. In the USW model, the signal phases for different antennas are accurately modeled \cite{usw1}, while the channel gains are still uniform as in the UPW model. More accurate than the USW model, the NUSW model appropriately captures the variations of both the phases and channel gains for different links across array elements \cite{haiyang_TWC}. However, it worth to note that all the three conventional models (TCMs) mentioned above, i.e., UPW, USW and NUSW, ignore the loss in channel gain caused by the effective antenna aperture \cite{zengyong_twc} and the polarization mismatch \cite{polarization_1,polarization_2}. The effective aperture denotes the projected antenna aperture that is orthogonal to the local wave propagation direction corresponding to the current element, and the polarization mismatch represents the angular difference in polarization between the local wave and the antenna \cite{NFC}. Consequently, the effective aperture and polarization loss vary across array elements. If such losses are neglected, the receive power can unlimitedly increase with the number of antennas and even exceed the transmit power, leading to the violation of the energy-conservation law \cite{NFC}.

Therefore, motivated by the aforementioned research gaps, we conduct a performance analysis for a near-field ISAC system based on an accurate channel model in this work. The main contributions of the paper are summarized as follows:

\begin{itemize}
 \item We propose a near-field ISAC framework for both downlink and uplink scenarios, where the base station (BS) is equipped with a uniform planar array (UPA). We employ channel model that is more accurate than the TCMs. In addition to precisely modelling the variations of signal phase and amplitude, this model takes into account the effects introduced by the effective antenna aperture and polarization mismatch for each element. 
 
\item We study the downlink near-field ISAC performance, considering three distinct scenarios: the sensing-centric (S-C) design, the communications-centric (C-C) design, and the Pareto optimal design. For each scenario, we derive SR, CR, and their high signal-to-noise ratio (SNR) slopes. To further validate the rationality of the model in comparison to the TCMs, we derive the asymptotic CRs and SRs for the arrays with an infinite number of elements. Furthermore, we consider upper bounds for the CRs and SRs under the scenario where the polarization mismatch is mitigated. Finally, we characterize the attainable SR-CR regions of downlink ISAC and FDSAC. 

\item We analyze the uplink near-field ISAC performance by considering the S-C design and the C-C design, with each employing different interference cancellation orders for S\&C signals at the BS. We derive the same metrics as in the downlink case and also obtain the achievable rate region of uplink ISAC by using the time-sharing strategy. 

\item Numerical results are presented, demonstrating that ISAC achieves a more extensive rate region than the FDSAC in both downlink and uplink cases. Additionally, when the number of array elements increases, the SRs and CRs of our model tend to finite limits, while those of the TCMs grow unlimitedly. For a given number of array elements, this performance gap narrows as the distances between the communication user (CU) and target to the BS increases.
\end{itemize}

\begin{figure}[!t]
    \centering
    \subfigbottomskip=5pt
	\subfigcapskip=0pt
\setlength{\abovecaptionskip}{0pt}
    \subfigure[A downlink/uplink ISAC system.]
    {
        \includegraphics[height=1.25in]{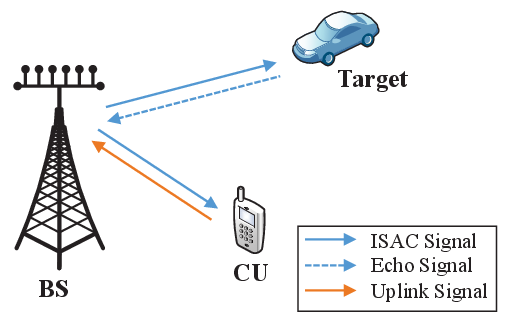}
	   \label{fig1}	
    }
    
    \vspace{5pt}
   \subfigure[System layout of the UPA.]
    {
        \includegraphics[height=0.33\textwidth]{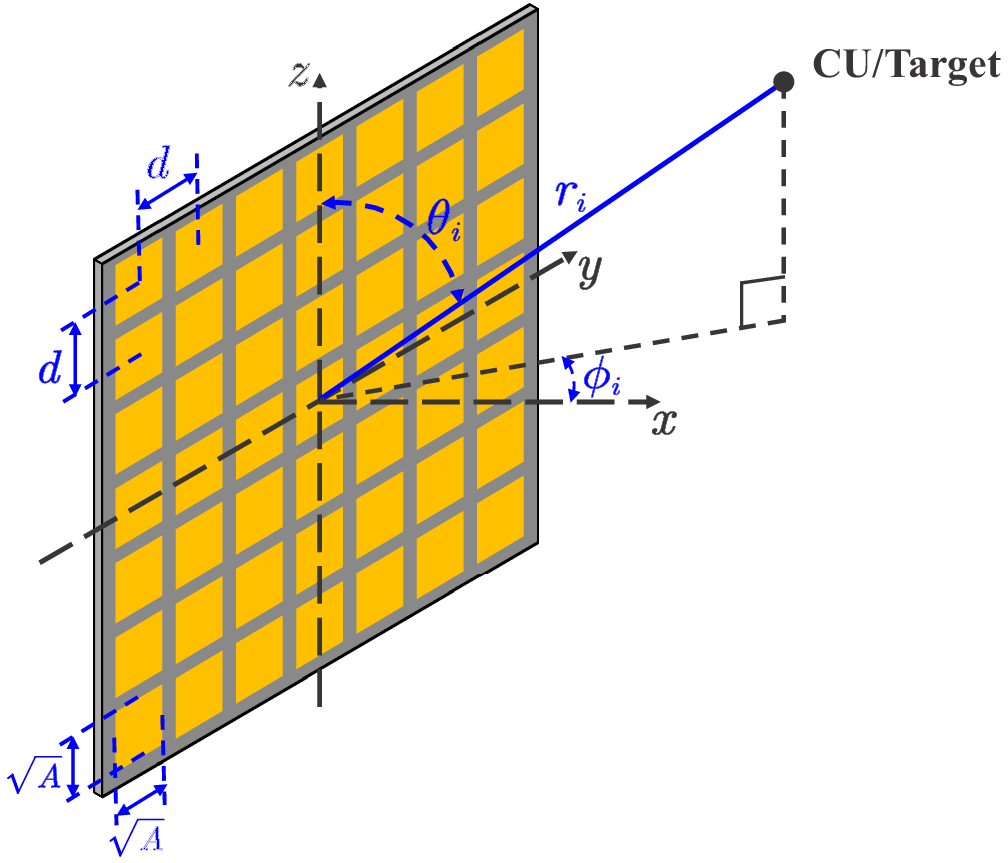}
	   \label{fig2}	
    }
\caption{Illustration of downlink/uplink near-field ISAC.}
    \label{model}
\end{figure}

The rest of this paper is organized as follows. In Section~\ref{system}, we introduce the conceptual framework of the ISAC model in the near-field region, addressing both downlink and uplink scenarios. Section~\ref{downlink} is dedicated to examining the near-field ISAC performance in the downlink case, presenting results of the three designs and the achievable SR-CR regions. Moving on to Section~\ref{uplink}, we explore the near-field ISAC performance in the uplink case, offering results for the two distinct designs and characterizing the rate region with the time-sharing strategy. Numerical results are present in Section~\ref{numerical}, and a concise conclusion is provided in Section~\ref{conclusion}.

\subsubsection*{Notations}
Throughout this paper, scalars, vectors, and matrices are denoted by non-bold, bold lower-case, and bold upper-case letters, respectively. For the vector $\mathbf{a}$, $[\mathbf{a}]_i$, ${\mathbf{a}}^{\mathsf{T}}$, ${\mathbf{a}}^{*}$, and ${\mathbf{a}}^{\mathsf{H}}$ denote the $i$th entry, transpose, conjugate, and conjugate transpose of $\mathbf{a}$, respectively. The notations $\lvert a\rvert$ and $\lVert \mathbf{a} \rVert$ denote the magnitude and norm of scalar $a$ and vector $\mathbf{a}$, respectively. The $N\times N$ identity matrix is denoted by ${\mathbf{I}}_N$. The set $\mathbbmss{C}$ stand for the complex spaces. The MI between random variables $X$ and $Y$ conditioned on $Z$ is denoted by $I\left(X;Y|Z\right)$. Finally, ${\mathcal{CN}}({\bm\mu},\mathbf{X})$ is used to denote the circularly-symmetric complex Gaussian distribution with mean $\bm\mu$ and covariance matrix $\mathbf{X}$.

\section{System Model}\label{system}
Consider a downlink/uplink near-field ISAC system, as illustrated in {\figurename} \ref{fig1}, where a dual-functional S\&C (DFSAC) BS equipped with an {$N$-antenna} UPA is communicating with a CU equipped with a single antenna, while concurrently performing sensing on a target. We consider a monostatic sensing setup at the BS. As shown in Fig.~\ref{fig2}, we assume that the UPA centered at the origin is deployed along the $y-z$ plane and $N=N_yN_z$ with $N_y$ and $N_z$ denoting the number of antennas along the $y$-axis and $z$-axis, respectively. The antenna size is denoted as $\sqrt{A}\times\sqrt{A}$, and the inter-element distance is represented as $d$ ($d>\sqrt{A}$). In particular, $\zeta \triangleq \frac{A}{d^2}\in \left( 0,1 \right] $ denotes the array occupation ratio (AOR) which measures the proportion of the entire UPA area occupied by antennas. Without loss of generality, we consider $N_y$ and $N_z$ as odd numbers. The central location of the $(n_y,n_z)$-th element is denoted as $\mathbf{p}_{n_y,n_z}=\left[ 0,n_yd,n_zd \right] ^\mathsf{T}$ for $n_y\in\left\{0,\pm 1, \ldots , \pm \frac{N_y-1}{2}\right\}$ and $n_z\in\left\{0,\pm 1, \ldots , \pm \frac{N_z-1}{2}\right\}$.

\vspace{-5pt}
\subsection{Near-field Channel Model}
As illustrated in Fig.~\ref{fig2}, we consider a CU/target located at a distance $r_i$ from the center of the UPA, with an elevation angle $\theta_i \in \left[ 0,\pi \right] $ and an azimuth angle $\phi_i \in \left[ -\frac{\pi}{2},\frac{\pi}{2} \right] $, where $i \in\left\{ \mathrm{c},\mathrm{s} \right\} $ with the subscripts ``$\mathrm{c}$'' and ``$\mathrm{s}$'' representing the CU and the target, respectively. Therefore, the location of the CU/target is given by
$\mathbf{r}_i=\left[ r_i\Psi_i ,r_i\Phi_i ,r_i\Omega_i \right] ^\mathsf{T}$, where $\Psi_i \triangleq \sin \theta_i \cos \phi_i $, $\Phi_i \triangleq \sin \theta_i \sin \phi_i $, and $\Omega_i \triangleq \cos \theta_i $. Accordingly, the distance between the CU/target and the $(n_y,n_z)$-th antenna is given as follows:
\begin{align}
r_{n_y,n_z}&=\left\| \mathbf{r}_i-\mathbf{p}_{n_y,n_z} \right\| \notag\\
&=r_i\sqrt{\left( n_y\epsilon_i -\Phi_i \right) ^2+\left( n_z\epsilon_i -\Omega_i \right) ^2+\Psi_i ^2},
\end{align}
where $\epsilon_i=\frac{d}{r_i}$. Particularly, since the inter-element distance $d$ is typically on the order of wavelength, we have $r_i\gg d$ and thus $\epsilon_i \ll 1$.

\begin{figure*}[!b]
\hrulefill
\begin{align}\label{channel_model}
h_{n_y,n_z}\left( r_i,\theta _i,\phi _i \right) &=\sqrt{A\underset{\text{{\small Free-space loss}}}{\underbrace{\frac{1}{4\pi r_{n_y,n_z}^{2}}}}\times \underset{\text{\small Effective aperture loss}}{\underbrace{\frac{r_i\Psi _i}{r_{n_y,n_z}}}}\times \underset{\text{\small Polarization loss}}{\underbrace{\frac{r_{i}^{2}\Psi _{i}^{2}+r_i\Psi _i\left( r_i\Omega _i-n_zd \right) ^2}{r_{n_y,n_z}^{2}}}}}{\rm{e}}^{{-\rm{j}}\frac{2\pi}{\lambda}r_{n_y,n_z}}\notag\\
&=\sqrt{A\frac{r_{i}^{3}\Psi _{i}^{3}+r_i\Psi _i\left( r_i\Omega _i-n_zd \right) ^2}{4\pi r_{n_y,n_z}^{5}}}{\rm{e}}^{{-\rm{j}}\frac{2\pi}{\lambda}r_{n_y,n_z}}.\tag{7}
\end{align}
\end{figure*}

Since the signals transmitted by different antennas are observed by the receiver from distinct angles, the resulting effective antenna aperture and polarization mismatch varies across the array. The effective antenna aperture is determined by the projection of the array normal to the direction of the signal, and the polarization mismatch is characterized by the squared norm of the inner product between the polarization vectors of the receiving mode and the transmitting mode, with the resulting loss in channel gain referred to as effective aperture loss and polarization loss, respectively \cite{NFC}. Consequently, under free-space line-of-sight propagation, the channel power gain between the $(n_y,n_z)$-th element and the CU/target is given by 
\begin{align}\label{UPA_model}
\!\!\left| h_{n_y,n_z}\!\left( \mathbf{r}_i \right) \right|^2\!=\!\!\int_{\mathcal{P} _{n_y},_{n_z}}\!{\mathcal{L} _1\left( \mathbf{r}_i,\mathbf{p} \right) \mathcal{L} _2\left( \mathbf{r}_i,\mathbf{p} \right) \mathcal{L} _3\left( \mathbf{r}_i,\mathbf{p} \right) d\mathbf{p}},  
\end{align}
where
\begin{align}
\mathcal{L} _1\left( \mathbf{r}_i,\mathbf{p} \right) &=\frac{1}{4\pi \left\| \mathbf{r}_i-\mathbf{p} \right\| ^2},\\
\mathcal{L} _2\left( \mathbf{r}_i,\mathbf{p} \right) &=\frac{\left( \mathbf{r}_i-\mathbf{p} \right) ^{\mathsf{T}}\hat{\mathbf{u}}_x}{\left\| \mathbf{r}_i-\mathbf{p} \right\|},\\
\mathcal{L} _3\left( \mathbf{r}_i,\mathbf{p} \right) &=\frac{\left| \boldsymbol{\rho }^{\mathsf{T}}\left( \mathbf{r}_i \right) \mathbf{e}\left( \mathbf{p},\mathbf{r}_i \right) \right|^2}{\left\| \mathbf{e}\left( \mathbf{p},\mathbf{r}_i \right) \right\| ^2}
\end{align}
denotes the free-space path loss, the effective aperture loss and the polarization loss, respectively. Furthermore, $\mathcal{P} _{n_y,n_z}=\left[ n_yd-\frac{\sqrt{A}}{2},n_yd+\frac{\sqrt{A}}{2} \right] \times \left[ n_zd-\frac{\sqrt{A}}{2},n_zd+\frac{\sqrt{A}}{2} \right] $ denotes the surface region of the $(n_y,n_z)$-th array element, $\mathbf{p}\in \mathcal{P} _{n_y,n_z}$ denotes the point located in this element, $\hat{\mathbf{u}}_x=\left[ 1,0,0 \right] $ represents the normal vector to the UPA, $\bm{\rho}(\mathbf{r}_i)$ denotes the normalized polarization vector at the CU/target, and $\mathbf{e}\left( \mathbf{p},\mathbf{r}_i \right) =\left( \mathbf{I}_3-\frac{\left( \mathbf{r}_i-\mathbf{p} \right) \left( \mathbf{r}_i-\mathbf{p} \right) ^{\mathsf{T}}}{\left\| \mathbf{r}_i-\mathbf{p} \right\| ^2} \right) \hat{\mathbf{J}}\left( \mathbf{p} \right) $ with $\hat{\mathbf{J}}\left( \mathbf{p} \right)$ being the normalized electric current vector at point $\mathbf{p}$ \cite{NFC}. Given that the size of each individual element is in the order of wavelength in practice, significantly smaller than $r_i$, we can assert that $r_i\gg \sqrt{A}$. Therefore, the variation of the channel coefficient among different points $\mathbf{p}\in \mathcal{P} _{n_y,n_z}$ can be considered negligible. As a result, \eqref{UPA_model} can be rewritten as
\begin{align}
\left| h_{n_y,n_z}\left( \mathbf{r}_i \right) \right|^2=A\prod_{k=1}^3{\mathcal{L} _k\left( \mathbf{r}_i,\mathbf{p}_{n_y,n_z} \right)}.
\end{align}

For mathematical tractability and to gain insights {into} the fundamental performance, in this paper, we consider a simplified case when the polarization vector at the CU/target and the electric current induced in the UPA both align in the $y$ direction, i.e., $\bm{\rho}(\mathbf{r}-i)=\hat{\mathbf{J}}\left( \mathbf{p} \right) =\left[ 0,1,0 \right] ^{\mathsf{T}}$. As a result, the near-field channel response between the $(n_y,n_z)$-th element and the CU/target can be obtained as in \eqref{channel_model}, shown as the bottom of this page.

To provide an intuitive comparison, we also briefly discuss the TCMs as follows. 
\subsubsection{UPW}
\setcounter{equation}{7}
\begin{align}
h_{n_y,n_z}^{\mathrm{UPW}}\left( r_i,\theta_i ,\phi_i \right) =\sqrt{\frac{A}{{4\pi r_i^2}}}{\rm{e}}^{{-\rm{j}}\frac{2\pi}{\lambda}\left( r_i-n_yd\Phi_i -n_zd\Omega_i \right)}.    
\end{align}
The UPW model is the commonly used model in far-field scenario, where both free-space path loss and the angles of the links between any element and the CU/target are assumed to be identical.

\subsubsection{USW}
\begin{align}
h_{n_y,n_z}^{\mathrm{USW}}\left( r_i,\theta _i,\phi_i \right) =\sqrt{\frac{A}{{4\pi r_i^2}}}{\rm{e}}^{{-\rm{j}}\frac{2\pi}{\lambda}r_{n_y,n_z}}.    
\end{align}
In the USW model, instead of using the planar wave, the USW is applied, accurately modelling the phase, while the path loss remains uniform. 

\subsubsection{NUSW}
\begin{align}
h_{n_y,n_z}^{\mathrm{NUSW}}\left( r_i,\theta_i ,\phi_i \right) =\sqrt{\frac{A}{{4\pi r_{n_y,n_z}^{2}}}}{\rm{e}}^{{-\rm{j}}\frac{2\pi}{\lambda}r_{n_y,n_z}}.    
\end{align}
Unlike the USW model, the NUSW model calculates the path loss for each link separately. 

It is important to note that none of the TCMs take into account the effective aperture loss and polarization loss. As we will see later, this ignorance can cause unreasonable consequence. Therefore, throughout this study, we investigate the performance of near-field ISAC based on the more accurate model given in \eqref{channel_model}.

\subsection{Downlink Signal Model}
Consider a signal matrix for the DFSAC denoted as $\mathbf{X}=\left[{\mathbf x}_1 \ldots {\mathbf x}_L\right]\in{\mathbbmss{C}}^{N\times L}$ transmitted from the BS, where $L$ denotes the duration of the sensing pulse/communication frame. For sensing purpose, each ${\mathbf x}_l\in{\mathbbmss{C}}^{N\times1}$ for $l=1,\ldots,L$ corresponds to the snapshot used for sensing during the $l$th time slot. In the context of communication, ${\mathbf x}_l$ represents the $l$th data symbol vector. Under the framework of multiple-input single-output (MISO) ISAC, we can write the ISAC signal $\mathbf{X}$ as follows:
\setcounter{equation}{11}
\begin{align}\label{dual_function_signal_matrix}
\mathbf{X}=\sqrt{p}\mathbf{w} \mathbf{s}^{\mathsf{H}} ,
\end{align}
where $\mathbf{w}\in \mathbbmss{C} ^{N\times 1}$ denotes the normalized beamforming vector with $\left\| \mathbf{w} \right\| ^2=1$, $p$ denotes the transmit power, and $\mathbf{s}=\left[{s}_1 \ldots {s}_L\right]^\mathsf{H}\in{\mathbbmss{C}}^{L\times 1}$ denotes the unit-power data streams intended for the CU with $L^{-1}\left\| \mathbf{s} \right\| ^2=1$.

\subsubsection{Communication Model}
The signal received at the CU is given by
\begin{align}
\mathbf{y}^{\mathsf{T}}_{\mathrm{c}}=\mathbf{h}_{\mathrm{c}}^{\mathsf{T}}\mathbf{X}+\mathbf{n}_{\mathrm{c}}^{\mathsf{H}}=\sqrt{p}\mathbf{h}_{\mathrm{c}}^{\mathsf{T}}\mathbf{ws}^{\mathsf{H}}+\mathbf{n}_{\mathrm{c}}^{\mathsf{H}},
\end{align}
where $\mathbf{h}_{\mathrm{c}}\in \mathbbmss{C}^{N\times 1}$ obtained by consolidating $\{h_{n_y,n_z}\left( r_{\mathrm{c}},\theta _{\mathrm{c}},\phi _{\mathrm{c}} \right)\}_{\forall n_y,n_z}$ into a vector represents the communication channel, and $\mathbf{n}_{\mathrm{c}}\in \mathbbmss{C}^{L\times 1}$ denotes the additive white Gaussian noise (AWGN) vector with each entry having zero mean and unit variance. Accordingly, the downlink CR is given by 
\begin{align}
 \mathcal{R} _{\mathrm{d},\mathrm{c}}=\log _2\left( 1+p\left| \mathbf{h}_{\mathrm{c}}^{\mathsf{T}}\mathbf{w}\right|^2 \right).   
\end{align}

\subsubsection{Sensing Model}
The received echo signal at the BS for target sensing is given by
\begin{align}\label{reflected_echo_signal_matrix}
{\mathbf{Y}}_{\mathrm{s}}={\mathbf{G}}{\mathbf{X}}+{\mathbf{N}}_{\mathrm{s}},
\end{align}
where $\mathbf{G}\in{\mathbbmss{C}}^{N\times N}$ represents the target response matrix, and ${\mathbf{N}}_{\mathrm{s}}\in{\mathbbmss{C}}^{N\times L}$ denotes the AWGN matrix with each entry having zero mean and unit variance. Specifically, the target response matrix can be modeled with the round-trip channel:
\begin{align}\label{G_model}
\mathbf{G}=\beta \mathbf{h}_{\mathrm{s}}\mathbf{h}_{\mathrm{s}}^{\mathsf{T}},
\end{align}
where $\beta\sim{\mathcal{CN}}\left(0,\alpha _s\right)$ denotes the complex amplitude of the target with the average strength of $\alpha _s$, and $\mathbf{h}_{\mathrm{s}}=[h_{n_y,n_z}\left( r_{\mathrm{s}},\theta _{\mathrm{s}},\phi _{\mathrm{s}} \right)]_{\forall n_y,n_z}\in \mathbbmss{C}^{N\times 1}$ represents the sensing link between the UPA and the target. Therefore, we can rewrite the reflected echo signal as
\begin{align}\label{echo}
\mathbf{Y}_{\mathrm{s}}=\sqrt{p}\beta \mathbf{h}_{\mathrm{s}}\mathbf{h}_{\mathrm{s}}^{\mathsf{T}}\mathbf{ws}^{\mathsf{H}}+\mathbf{N}_{\mathrm{s}}.
\end{align}

In this study, we assume that the target's position is accurately tracked and known in advance by the BS. Therefore, our focus lies in estimating the reflection coefficient $\beta$. The sensing task involves extracting environmental information contained in $\beta$ from the reflected echo signal ${\mathbf{Y}}_{\mathrm{s}}$, with the knowledge of $\mathbf{X}$. {Information-theoretic} bounds for this sensing task are quantified by the sensing mutual information (MI), which denotes the MI between ${\mathbf{Y}}_{\mathrm{s}}$ and $\beta$, conditioned on the ISAC signal $\mathbf{X}$ \cite{LiuAn}. In evaluating the sensing performance, we utilize the SR as the performance metric, defined as the sensing MI per unit time \cite{Ouyang2022_WCL,Tang2019_TSP,Zhang2021_JSTSP}. Assuming that each DFSAC symbol lasts $1$ unit time, the SR is expressed as
\begin{align}\label{SR_define}
\mathcal{R} _{\mathrm{d},\mathrm{s}}=L^{-1}I\left( \mathbf{Y}_{\mathrm{s}};\beta |\mathbf{X} \right) ,    
\end{align}
In particular, $\mathcal{R} _{\mathrm{d},\mathrm{s}}$ can be calculated as follows.
\begin{lemma}\label{SR_lemma}
For a given $\mathbf{w}$, the downlink SR is given by
\begin{align}
\mathcal{R} _{\mathrm{d},\mathrm{s}}=\frac{1}{L}\log _2\left( 1+pL\alpha _{\mathrm{s}}\left\| \mathbf{h}_{\mathrm{s}} \right\| ^2\left| \mathbf{h}_{\mathrm{s}}^{\mathsf{T}}\mathbf{w} \right|^2 \right) .
\end{align}
\end{lemma}
\begin{IEEEproof}
Please refer to Appendix~\ref{Appendix:A}.   
\end{IEEEproof}
With the proposed downlink near-field ISAC framework, our objective is to assess its S\&C performance by examining the CR and SR, both of which are influenced by the beamforming vector $\mathbf{w}$. However, finding an optimal $\mathbf{w}$ that that can effectively enhance both $\mathcal{R}_{\mathrm{d},\mathrm{c}}$ and $\mathcal{R}_{\mathrm{d},\mathrm{s}}$ concurrently presents a formidable challenge. In order to tackle this issue, we present three distinctive scenarios within the downlink near-field ISAC framework in Section~\ref{downlink}. The first scenario is referred as the C-C design, with the primary aim of optimizing the CR. In the second scenario, we delve into the S-C design, which seeks to maximize the SR. Finally, we focus on the Pareto optimal design, aiming to identify the Pareto boundary of the achievable SR-CR region.

\subsection{Uplink Signal Model}
In the uplink case, the signal observed by the BS reads
\begin{align} \label{BS_receive}
\mathbf{Y}=\sqrt{p_{\mathrm{c}}}\mathbf{h}_{\mathrm{c}}\mathbf{s}_{\mathrm{c}}^\mathsf{H}+\sqrt{p_{\mathrm{s}}}\mathbf{Gws}_{\mathrm{s}}^\mathsf{H}+\mathbf{N}_{\mathrm{u}},
\end{align}
where $p_{\mathrm{c}}$ denotes the communication power,  $\mathbf{s}_{\mathrm{c}}=\left[ s_{\mathrm{c},1}\ldots s_{\mathrm{c},L} \right]^\mathsf{H}\in{\mathbbmss{C}}^{ L\times 1}$ denotes the communication messages sent by the CU subject to $\mathbb{E} \left\{ \mathbf{s}_{\mathrm{c}}\mathbf{s}_{\mathrm{c}}^{\mathsf{H}} \right\} =\mathbf{I}_L$, $p_{\mathrm{s}}$ denotes the sensing power,  $\mathbf{s}_{\mathrm{s}}=\left[ s_{\mathrm{s},1}\ldots s_{\mathrm{s},L} \right]^\mathsf{H}\in{\mathbbmss{C}}^{ L\times 1}$ denotes the sensing pulse subject to $L^{-1}\lVert \mathbf{s}_{\mathrm{s}} \rVert^2=1$, and $\mathbf{N}_{\mathrm{u}}=\left[ \mathbf{n}_{\mathrm{u},1}\ldots \mathbf{n}_{\mathrm{u},L} \right]^\mathsf{H}\in{\mathbbmss{C}}^{N \times L}$ is the standard AWGN matrix. 

Upon reception of the signal described above, the BS faces the challenge of decoding both the communication signal and the environmental information contained in the target response matrix $\mathbf{G}$. To effectively manage this inter-functionality interference (IFI), we can employ successive interference cancellation (SIC) with two different decoding orders \cite{uplink_SIC}. In the first SIC order, the BS firstly identifies the target response signal, regarding the communication signal as interference. Then, the identified sensing signal will be removed from the superposed signal, leaving the remaining part for communication signal detection. In contrast, the second SIC order begins by detecting the communication signal, treating the echo signal as interference. Following this, the communication signal undergoes subtraction, thereby preserving the residual signal for the purpose of sensing. The primary observation is that the first SIC order excels in optimizing communication performance, whereas the second SIC order enhances sensing performance. As a result, we denote these two SIC orders as the C-C design and the S-C design, respectively, which will be investigated in Section~\ref{uplink}.

\section{Downlink Near-field ISAC} \label{downlink}
This section introduces three scenarios for downlink near-field ISAC: C-C design, S-C design, and Pareto optimal design. In each scenario, the SR, CR, and their asymptotic performance are investigated. Furthermore, performance upper bound in the absence of polarization loss is explored. Finally, the downlink SR-CR region achieved by near-field ISAC is characterized.
\subsection{Communications-Centric Design}
Under the C-C design, the beamforming vector $\mathbf{w}$ is set to maximize the downlink CR, which is given by
\begin{align}
\mathbf{w}_{\mathrm{c}}=\argmax\nolimits_{\mathbf{w}}\mathcal{R} _{\mathrm{d},\mathrm{c}}=\argmax\nolimits_{\mathbf{w}}\lvert\mathbf{h}_{\mathrm{c}}^{\mathsf{T}}\mathbf{w}\rvert=\frac{\mathbf{h}^{*}_{\mathrm{c}}}{\left\| \mathbf{h}_{\mathrm{c}} \right\|} .
\end{align}
With the beamforming vector $\mathbf{w}_{\mathrm{c}}$, we investigate the downlink communication performance and sensing performance under the C-C design, respectively, in the following subsections.

\begin{figure*}[!b]
\hrulefill
\begin{align}\label{delta}
\delta _i\left( y,z \right) =\frac{2}{3}\arctan \left( \frac{yz}{\Psi _i\sqrt{\Psi _{i}^{2}+y^2+z^2}} \right) +\frac{\Psi _iyz}{3\left( \Psi _{i}^{2}+y^2 \right) \sqrt{\Psi _{i}^{2}+y^2+z^2}}, \quad i\in \left\{ \mathrm{c},\mathrm{s} \right\} .\tag{24}
\end{align}
\end{figure*}

\subsubsection{Performance of Communications}
Given $\mathbf{w}=\mathbf{w}_{\mathrm{c}}$, the CR can be written as follows:
\begin{align}\label{eq_Rcc}
\mathcal{R} _{\mathrm{d},\mathrm{c}}^{\mathrm{c}}=\log _2\left( 1+p\left\| \mathbf{h}_{\mathrm{c}} \right\| ^2 \right) .
\end{align}
The following theorem provides an exact closed-form expression of $\mathcal{R} _{\mathrm{d},\mathrm{c}}^{\mathrm{c}}$ and its high-SNR approximation.
\vspace{-5pt}
\begin{theorem}\label{CC_CR_theorem}
Under the C-C design, the downlink CR {is given by}
\begin{align}\label{CC_CR}
\mathcal{R} _{\mathrm{d},\mathrm{c}}^{\mathrm{c}}=\log _2\left(1+\frac{p\zeta}{4\pi}\sum_{y\in \mathcal{Y} _{\mathrm{c}}}{\sum_{z\in \mathcal{Z} _{\mathrm{c}}}{\delta _{\mathrm{c}}\left( y,z \right)}} \right)  ,  
\end{align}
where $\mathcal{Y} _{\mathrm{c}}=\left\{ \frac{N_y\epsilon _{\mathrm{c}}}{2}\pm \Phi_{\mathrm{c}} \right\} $, $\mathcal{Z} _{\mathrm{c}}=\left\{ \frac{N_z\epsilon _{\mathrm{c}}}{2}\pm \Omega_{\mathrm{c}} \right\} $, and $\delta _{\mathrm{c}}\left( y,z \right) $ is defined in \eqref{delta}, shown as the bottom of this page. {For large $p$ (i.e., high-SNR), we have}
\setcounter{equation}{24}
\begin{align}\label{CC_CR_p}
{\mathcal{R}_{\mathrm{d},\mathrm{c}}^{\mathrm{c}}
\!\approx}\log _2p+\log _2\!\left(\frac{\zeta}{4\pi}\sum_{y\in \mathcal{Y} _{\mathrm{c}}}\!{\sum_{z\in \mathcal{Z} _{\mathrm{c}}}{\!\delta _{\mathrm{c}}\left( y,z \right)}}\!\right)\!.
\end{align}
\end{theorem}
\vspace{-5pt}
\begin{IEEEproof}
Please refer to Appendix \ref{Appendix:B}.
\end{IEEEproof}
\vspace{-5pt}
\begin{remark}
\eqref{CC_CR_p} indicates that the high-SNR slope and the high-SNR power offset of $\mathcal{R} _{\mathrm{d},\mathrm{c}}^{\mathrm{c}}$ are, respectively, given by $\mathcal{S} _{\mathrm{d},\mathrm{c}}^{\mathrm{c}}=1$ and $\mathcal{O} _{\mathrm{d},\mathrm{c}}^{\mathrm{c}}=\log _2\left(\frac{\zeta}{4\pi}\sum_{y\in \mathcal{Y} _{\mathrm{c}}}{\sum_{z\in \mathcal{Z} _{\mathrm{c}}}{\delta _{\mathrm{c}}\left( y,z \right)}}\right)$.
\end{remark}

To gain further insights for the CR under the near-field channel model, we next investigate its asymptotic behaviour when the UPA has an infinite number of elements, i.e., $N_y,N_z\rightarrow \infty$.
\vspace{-5pt}
\begin{corollary} \label{CC_CR_M_cor}
As $N_y,N_z\!\rightarrow \!\infty$, the asymptotic CR {is given by}
\begin{align}\label{CC_CR_M}
\lim_{N_y,N_z\rightarrow \infty}\mathcal{R} _{\mathrm{d},\mathrm{c}}^{\mathrm{c}}=\log _2\left( 1+\frac{p\zeta}{3 } \right) .
\end{align}
\end{corollary}
\vspace{-5pt}
\begin{IEEEproof}
Since $\lim_{y,z\rightarrow \infty} \arctan \left( \frac{yz}{\Psi _{\mathrm{c}}\sqrt{\Psi _{\mathrm{c}}^{2}+y^2+z^2}} \right) =\frac{\pi}{2}$ and $\lim_{y,z\rightarrow \infty} \frac{\Psi _{\mathrm{c}}yz}{3\left( \Psi _{\mathrm{c}}^{2}+y^2 \right) \sqrt{\Psi _{\mathrm{c}}^{2}+y^2+z^2}}=0$, we have 
$\lim_{N_y,N_z\rightarrow \infty} \delta _{\mathrm{c}}\left( y,z \right)  =\frac{\pi}{3}$ for $y\in\mathcal{Y}_{\mathrm{c}}$ and $z\in\mathcal{Z}_{\mathrm{c}}$. 
\end{IEEEproof}

\begin{remark}
The results in \textbf{Corollary}~\ref{CC_CR_M_cor} indicate that as $N_y,N_z\rightarrow \infty $, {rather than growing unboundedly, the downlink CR in the C-C design converges to a finite quantity that is increasing in the AOR.}
\end{remark}

Notably, in traditional antenna design, there are methods to mitigate or avoid polarization mismatch, such as polarization matching between the transmitter and receiver \cite{polar_match}, multi-polarized antennas \cite{multi_polar}, and adaptive polarization \cite{adapt_polar}. Accordingly, we examine a scenario where the polarization mismatch is considered to be avoided, i.e., $\mathcal{L} _3\left( \mathbf{r}_i,\mathbf{p} \right) =1$, establishing an ideal upper bound for system performance. Under this consideration, the near-field channel model in \eqref{channel_model} can be rewritten as
\begin{align}
\bar{h}_{n_y,n_z}\left( r_i,\theta _i,\phi _i \right) =\sqrt{\frac{Ar_i\Psi _i}{4\pi r_{n_y,n_z}^{3}}}{\rm{e}}^{{-\rm{j}}\frac{2\pi}{\lambda}r_{n_y,n_z}}.   
\end{align}
Consolidating $\{\bar{h}_{n_y,n_z}\!\left( r_{i},\theta _{i},\phi _{i} \right)\}_{\forall n_y,n_z}\!$ into a vector gives $\bar{\mathbf{h}}_{i}$.
\begin{corollary}\label{rcc_polar_cor}
When the polarization loss is eliminated, the downlink CR of the C-C design is given by 
\begin{align}\label{rcc_polar}
\bar{\mathcal{R}} _{\mathrm{d},\mathrm{c}}^{\mathrm{c}}\!=\!\log _2\!\Bigg(\!1\!+\!\frac{p\zeta}{4\pi}\sum_{y\in \mathcal{Y} _{\mathrm{c}}}{\sum_{z\in \mathcal{Z} _{\mathrm{c}}}\!{\arctan \!\bigg( \frac{yz}{\Psi _{\mathrm{c}}\sqrt{\Psi _{\mathrm{c}}^{2}\!+\!y^2\!+\!z^2}} \bigg)}} \!\Bigg).
\end{align}
When $N_y,N_z\rightarrow \infty $, we have
\begin{align}
\lim_{N_y,N_z\rightarrow \infty}\bar{\mathcal{R}} _{\mathrm{d},\mathrm{c}}^{\mathrm{c}}=\log _2\left( 1+\frac{p\zeta}{2 } \right),   
\end{align}
which is a finite value larger than $\lim_{N_y,N_z\rightarrow \infty}\mathcal{R} _{\mathrm{d},\mathrm{c}}^{\mathrm{c}}$.
\end{corollary}
\begin{IEEEproof}
Please refer to Appendix~\ref{Appendix:ex}.
\end{IEEEproof}

\subsubsection{Performance of Sensing}
When the beamforming vector $\mathbf{w}_{\mathrm{c}}$ is applied, the downlink SR under the C-C design reads
\begin{align}\label{CC_SR}
\mathcal{R}_{\mathrm{d},\mathrm{s}}^{\mathrm{c}}=\frac{1}{L}\log _2\left( 1+pL\alpha _{\mathrm{s}}\left\| \mathbf{h}_{\mathrm{s}} \right\| ^2\frac{\left| \mathbf{h}_{\mathrm{s}}^{\mathsf{T}} \mathbf{h}^{*}_{\mathrm{c}}\right|^2}{\left\| \mathbf{h}_{\mathrm{c}} \right\| ^2} \right).     
\end{align}
For clarity, we define the inner product of the normalized channels, namely
\begin{align}
\rho \triangleq\frac{\left| \mathbf{h}_{\mathrm{c}}^{\mathsf{H}}\mathbf{h}_{\mathrm{s}} \right|^2}{\left\| \mathbf{h}_{\mathrm{c}} \right\|^2 \left\| \mathbf{h}_{\mathrm{s}} \right\|^2}\in \left[ 0,1 \right],    
\end{align}
as the channel correlation factor (CCF), which measures the correlation between the communication and sensing channels. Accordingly, \eqref{CC_SR} can be rewritten as
\begin{align}
\mathcal{R} _{\mathrm{d},\mathrm{s}}^{\mathrm{c}} =\frac{1}{L}\log _2\left( 1+pL\alpha _{\mathrm{s}}\rho \left\| \mathbf{h}_{\mathrm{s}} \right\| ^4 \right),     
\end{align}
leading to the derivation of the following theorem.

\begin{figure} [t!]
\centering
\includegraphics[height=0.27\textwidth]{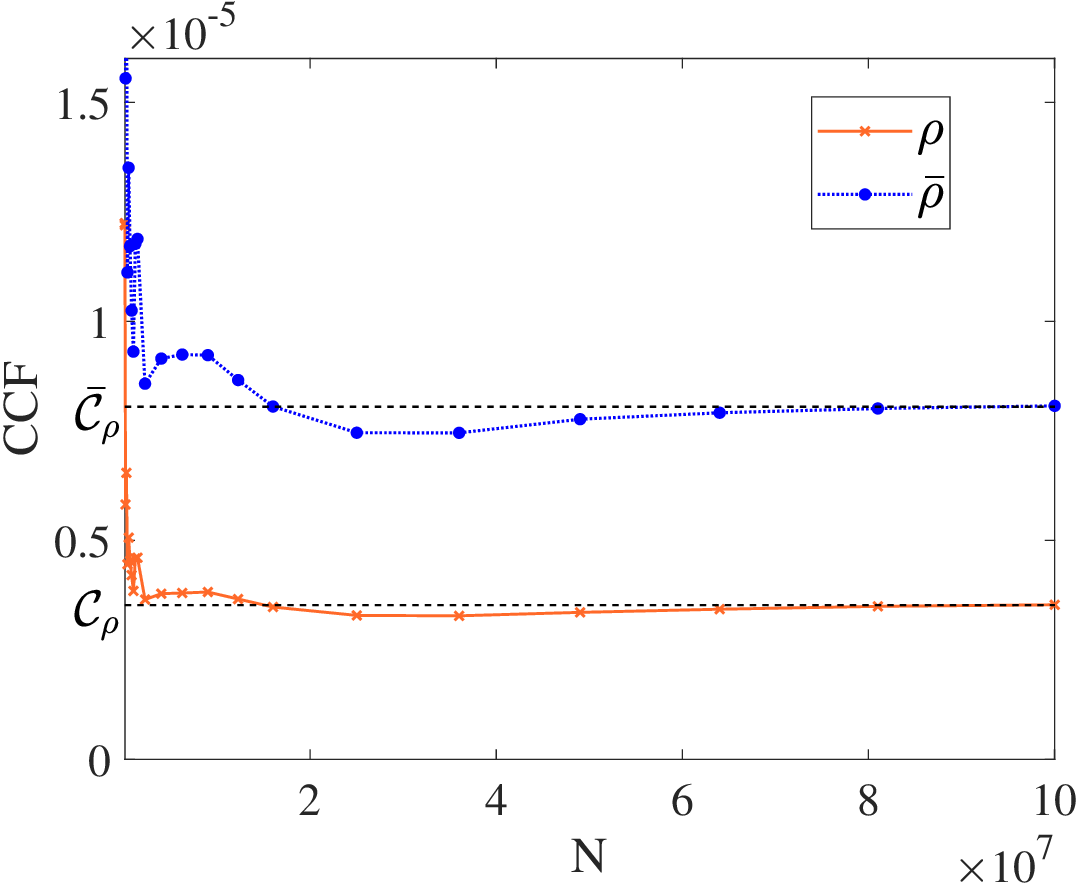}
 \caption{CCF versus $N$.}
 \vspace{-5pt}
 \label{figrho}
\end{figure}

\begin{theorem}\label{cc_SR_theorem}
The SR under the C-C design can be expressed as
\begin{align}
\mathcal{R} _{\mathrm{d},\mathrm{s}}^{\mathrm{c}}\!=\!\frac{1}{L}\log _2\!\left[ 1\!+\!\frac{pL\alpha _{\mathrm{s}}\zeta ^2\rho }{16\pi ^2}\!\left( \sum_{y\in \mathcal{Y} _{\mathrm{s}}}\!{\sum_{z\in \mathcal{Z} _{\mathrm{s}}}\!{\delta _{\mathrm{s}}\left( y,z \right)}} \right) ^2 \right] \!,
\end{align}
where $\mathcal{Y} _{\mathrm{s}}=\left\{ \frac{N_y\epsilon _{\mathrm{s}}}{2}\pm \Phi_{\mathrm{s}} \right\} $, $\mathcal{Z} _{\mathrm{s}}=\left\{ \frac{N_z\epsilon _{\mathrm{s}}}{2}\pm \Omega_{\mathrm{s}} \right\} $, and $\delta _{\mathrm{s}}\left( y,z \right) $ is defined in \eqref{delta}. {For large $p$, its high-SNR approximation is given by}
\begin{align}\label{CC_SR_p}
{\mathcal{R} _{\mathrm{d},\mathrm{s}}^{\mathrm{c}}\approx}\frac{1}{L}\!\left[\log _2p \!+\!2\log _2\!\left( \frac{\zeta\sqrt{L\alpha _{\mathrm{s}}\rho} }{4\pi}\sum_{y\in \mathcal{Y} _{\mathrm{s}}}{\sum_{z\in \mathcal{Z} _{\mathrm{s}}}{\delta _{\mathrm{s}}\left( y,z \right)}} \right)\right] .     
\end{align}
\end{theorem}
\vspace{-5pt}
\begin{IEEEproof}
The proof of this result is similar to that of \textbf{Theorem~\ref{CC_CR_theorem}}.
\end{IEEEproof}
\vspace{-5pt}
\begin{remark}
The results in \eqref{CC_SR_p} indicate that the high-SNR slope and the high-SNR power offset of $\mathcal{R} _{\mathrm{d},\mathrm{s}}^{\mathrm{c}}$ are, respectively, given by $\mathcal{S} _{\mathrm{d},\mathrm{s}}^{\mathrm{c}}=\frac{1}{L}$ and $\mathcal{O} _{\mathrm{d},\mathrm{s}}^{\mathrm{c}}=2\log _2\left( \frac{\zeta\sqrt{L\alpha _{\mathrm{s}}\rho } }{4\pi}\sum_{y\in \mathcal{Y} _{\mathrm{s}}}{\sum_{z\in \mathcal{Z} _{\mathrm{s}}}{\delta _{\mathrm{s}}\left( y,z \right)}} \right)$.
\end{remark}
While directly computing $\lim_{N_y,N_z\rightarrow \infty}\rho$ is intractable, we can infer that $\lim_{N_y,N_z\rightarrow \infty} \mathcal{R} _{\mathrm{d},\mathrm{s}}^{\mathrm{c}}\leqslant\frac{1}{L}\log _2\big( 1+\frac{pL\alpha _{\mathrm{s}}\zeta ^2}{9} \big) $, as $\rho\leqslant 1$. This suggests that $\lim_{N_y,N_z\rightarrow \infty} \mathcal{R} _{\mathrm{d},\mathrm{s}}^{\mathrm{c}}$ is a finite value. To reinforce this observation, we plot $\rho$ in terms of the number of antennas in {\figurename} \ref{figrho}, which is generated by averaging the CCFs across $10^4$ channel realizations. As can be observed from this graph, the CCF converges to some constant (denoted as $\mathcal{C}_\rho$) as $N_y,N_z\rightarrow \infty$. Based on this analysis, we derive the following corollary.
\begin{corollary} \label{CC_SR_M_cor}
When $N_y,N_z\rightarrow \infty $, the asymptotic SR under the C-C design is given by
\begin{align} 
\lim_{N_y,N_z\rightarrow \infty} \mathcal{R} _{\mathrm{d},\mathrm{s}}^{\mathrm{c}}=\frac{1}{L}\log _2\left( 1+\frac{\mathcal{C}_{\rho}pL\alpha _{\mathrm{s}}\zeta ^2}{9} \right) . \label{SC_SR_M} 
\end{align}
\end{corollary}
\begin{IEEEproof}
The proof of this result is similar to that of \textbf{Corollary}~\ref{CC_CR_M_cor}.
\end{IEEEproof}

\begin{remark}
The results in \textbf{Corollary}~\ref{CC_SR_M_cor} indicates that as $N_y,N_z\rightarrow \infty $, rather than growing unboundedly, the SR of the downlink C-C design converges to a finite quantity that is increasing in the AOR.
\end{remark}

The downlink SR of the C-C design in the absence of polarization loss is derived in the following corollary.
\begin{corollary}\label{rcs_polar_cor}
Without polarization loss, the downlink SR of the C-C design is given by 
\begin{align}\label{rcs_polar}
\bar{\mathcal{R}} _{\mathrm{d},\mathrm{s}}^{\mathrm{c}}=&\log _2\Bigg[1+\frac{pL\alpha _{\mathrm{s}}\zeta ^2\bar{\rho} }{16\pi ^2}\times \notag\\
&\Bigg(\sum_{y\in \mathcal{Y} _{\mathrm{s}}}{\sum_{z\in \mathcal{Z} _{\mathrm{s}}}\!{\arctan \!\bigg( \frac{yz}{\Psi _{\mathrm{s}}\sqrt{\Psi _{\mathrm{s}}^{2}\!+\!y^2\!+\!z^2}} \bigg)}}\Bigg)^{\!2} ~\Bigg],
\end{align}
where $\bar{\rho} \triangleq\frac{\left| \bar{\mathbf{h}}_{\mathrm{c}}^{\mathsf{H}}\bar{\mathbf{h}}_{\mathrm{s}} \right|^2}{\left\| \bar{\mathbf{h}}_{\mathrm{c}} \right\|^2 \left\| \bar{\mathbf{h}}_{\mathrm{s}} \right\|^2}$.  When $N_y,N_z\rightarrow \infty $, we have
\begin{align}
\lim_{N_y,N_z\rightarrow \infty} \bar{\mathcal{R}} _{\mathrm{d},\mathrm{s}}^{\mathrm{c}}=\frac{1}{L}\log _2\left( 1+\frac{\bar{\mathcal{C}}_{\rho}pL\alpha _{\mathrm{s}}\zeta ^2}{4} \right) ,   
\end{align}
which is a finite value with $\bar{\mathcal{C}}_{\rho}=\lim_{N_y,N_z\rightarrow \infty} \bar{\rho}$, as depicted in {\figurename} \ref{figrho}.
\end{corollary}
\begin{IEEEproof}
The proof of this result is similar to that of \textbf{Corollary~\ref{rcc_polar_cor}}.
\end{IEEEproof}

\subsection{Sensing-Centric Design}
In this subsection, we investigate the downlink S-C design. Under the S-C design, the beamforming vector is set to maximize the SR, which is given by
\begin{align}
\mathbf{w}_{\mathrm{s}}=\argmax\nolimits_{\mathbf{w}}\mathcal{R} _{\mathrm{d},\mathrm{s}}=\argmax\nolimits_{\mathbf{w}}\lvert\mathbf{h}_{\mathrm{s}}^{\mathsf{T}}\mathbf{w}\rvert=\frac{\mathbf{h}^{*}_{\mathrm{s}}}{\left\| \mathbf{h}_{\mathrm{s}} \right\|}.
\end{align}

\subsubsection{Performance of Sensing}
Given $\mathbf{w}=\mathbf{w}_{\mathrm{s}}$, the downlink SR can be written as follows:
\begin{align}
\mathcal{R}_{\mathrm{d},\mathrm{s}}^{\mathrm{s}}=\frac{1}{L}\log _2\left( 1+pL\alpha _{\mathrm{s}}\left\| \mathbf{h}_{\mathrm{s}} \right\| ^4 \right) .
\end{align}
The following theorem provides an exact closed-form expression of $\mathcal{R} _{\mathrm{d},\mathrm{s}}^{\mathrm{s}}$ and its high-SNR approximation.
\begin{theorem}\label{SC_SR_theorem}
The SR achieved by the S-C design {is given by}
\begin{align}\label{SC_SR}
\mathcal{R} _{\mathrm{d},\mathrm{s}}^{\mathrm{s}}\!=\!\frac{1}{L}\log _2\!\left[ 1\!+\!\frac{pL\alpha _{\mathrm{s}}\zeta ^2}{16\pi ^2}\!\left( \sum_{y\in \mathcal{Y} _{\mathrm{s}}}{\sum_{z\in \mathcal{Z} _{\mathrm{s}}}\!{\delta _{\mathrm{s}}\left( y,z \right)}} \right)^2 \right]  .
\end{align}
{For large $p$, we have }
\begin{align}\label{SC_SR_p}
{\mathcal{R} _{\mathrm{d},\mathrm{s}}^{\mathrm{s}}\!\approx}\frac{1}{L}\!\left[\log _2p\!+\!2\log _2\!\left( \!\frac{\sqrt{L\alpha _{\mathrm{s}}}\zeta}{4\pi}\!\sum_{y\in \mathcal{Y} _{\mathrm{s}}}\!{\sum_{z\in \mathcal{Z} _{\mathrm{s}}}\!{\delta _{\mathrm{s}}\!\left( y,z \right)}} \!\right)\!\right]\!.
\end{align}
\end{theorem}
\vspace{-5pt}
\begin{IEEEproof}
The proof of this result is similar to that of \textbf{Theorem}~\ref{CC_CR_theorem}.
\end{IEEEproof}

\begin{remark}
\eqref{SC_SR_p} indicates that the high-SNR slope and the high-SNR power offset of $\mathcal{R} _{\mathrm{d},\mathrm{s}}^{\mathrm{s}}$ are, respectively, given by $\mathcal{S} _{\mathrm{d},\mathrm{s}}^{\mathrm{s}}=\frac{1}{L}$ and $\mathcal{O} _{\mathrm{d},\mathrm{s}}^{\mathrm{s}}=2\log _2\left( \frac{\sqrt{L\alpha _{\mathrm{s}}}\zeta}{4\pi}\sum_{y\in \mathcal{Y} _{\mathrm{s}}}{\sum_{z\in \mathcal{Z} _{\mathrm{s}}}{\delta _{\mathrm{s}}\left( y,z \right)}} \right)$.
\end{remark}

We also investigate the asymptotic SR under the C-C design when the UPA has an infinite number of elements.
\vspace{-5pt}
\begin{corollary} \label{SC_SR_M_cor}
When $N_y,N_z\rightarrow \infty $, the asymptotic SR under the S-C design is given by
\begin{align} 
\lim_{N_y,N_z\rightarrow \infty} \mathcal{R} _{\mathrm{d},\mathrm{s}}^{\mathrm{s}}\!=\!\frac{1}{L}\log _2\left( 1\!+\!\frac{pL\alpha _{\mathrm{s}}\zeta ^2}{9} \right) .
\end{align}
\end{corollary}
\begin{IEEEproof}
The proof of this result is similar to that of \textbf{Corollary}~\ref{CC_CR_M_cor}.
\end{IEEEproof}

\begin{remark}
The results in \textbf{Corollary}~\ref{SC_SR_M_cor} indicate that as $N_y,N_z\rightarrow \infty $, {rather than growing unboundedly, the SR of the downlink S-C design converges to a finite quantity that is increasing in the AOR.}
\end{remark}

When the polarization mismatch is avoided, the SR under the S-C design can also be improved. 
\begin{corollary}\label{rss_polar_cor}
Without polarization loss, the downlink SR of the S-C design is given by 
\begin{align}\label{rss_polar}
\bar{\mathcal{R}} _{\mathrm{d},\mathrm{s}}^{\mathrm{s}}=&\log _2\Bigg[1+\frac{pL\alpha _{\mathrm{s}}\zeta ^2}{16\pi ^2}\times \notag\\
&\Bigg(\sum_{y\in \mathcal{Y} _{\mathrm{s}}}{\sum_{z\in \mathcal{Z} _{\mathrm{s}}}\!{\arctan \!\bigg( \frac{yz}{\Psi _{\mathrm{s}}\sqrt{\Psi _{\mathrm{s}}^{2}\!+\!y^2\!+\!z^2}} \bigg)}}\Bigg)^{\!2} ~\Bigg].
\end{align}
When $N_y,N_z\rightarrow \infty $, we have
\begin{align}
\lim_{N_y,N_z\rightarrow \infty} \bar{\mathcal{R}} _{\mathrm{d},\mathrm{s}}^{\mathrm{s}}=\frac{1}{L}\log _2\left( 1+\frac{pL\alpha _{\mathrm{s}}\zeta ^2}{4} \right) ,   
\end{align}
which is a finite value larger than $\lim_{N_y,N_z\rightarrow \infty}\mathcal{R} _{\mathrm{d},\mathrm{s}}^{\mathrm{s}}$.
\end{corollary}
\begin{IEEEproof}
The proof of this result is similar to that of \textbf{Corollary~\ref{rcc_polar_cor}}.
\end{IEEEproof}

\subsubsection{Performance of Communications}
When the beamforming vector $\mathbf{w}_{\mathrm{s}}$ is applied, the downlink CR under the S-C design is given by
\begin{align} 
\mathcal{R} _{\mathrm{d},\mathrm{c}}^{\mathrm{s}}=\log _2\!\left( 1+p\frac{\left| \mathbf{h}_{\mathrm{c}}^{\mathsf{T}} \mathbf{h}^{*}_{\mathrm{s}} \right|^2}{\left\| \mathbf{h}_{\mathrm{s}} \right\| ^2} \right)=\log _2\!\left( 1\!+p{\left\| \mathbf{h}_{\mathrm{c}} \right\| ^2}\rho \right).
\end{align}
The following theorem provides {an expression} of $\mathcal{R} _{\mathrm{d},\mathrm{c}}^{\mathrm{s}}$ and its high-SNR approximation.

\begin{theorem}\label{SC_CR_theorem}
The downlink CR of the S-C design is given by
\begin{align}
\mathcal{R} _{\mathrm{d},\mathrm{c}}^{\mathrm{s}}=\log _2\left( 1+\frac{p\zeta \rho }{4\pi}\sum_{y\in \mathcal{Y}_{\mathrm{c}}}{\sum_{z\in \mathcal{Z}_{\mathrm{c}}}{\delta _{\mathrm{c}}\left( y,z \right)}} \right) .
\end{align}
{For large $p$, its high-SNR approximation is given by}
\begin{align}\label{SC_CR_p}
{\mathcal{R} _{\mathrm{d},\mathrm{c}}^{\mathrm{s}}\approx}\log _2p+\log _2\!\left( \frac{\zeta \rho }{4\pi}\sum_{y\in \mathcal{Y}_{\mathrm{c}}}{\sum_{z\in \mathcal{Z}_{\mathrm{c}}}{\delta _{\mathrm{c}}\left( y,z \right)}} \right)  .     
\end{align}
\end{theorem}
\begin{IEEEproof}
The proof of this result is similar to that of \textbf{Theorem}~\ref{cc_SR_theorem}.
\end{IEEEproof}

\begin{remark} 
\eqref{SC_CR_p} indicates that the high-SNR slope and the high-SNR power offset of $\mathcal{R} _{\mathrm{d},\mathrm{c}}^{\mathrm{s}}$ are, respectively, given by $\mathcal{S} _{\mathrm{d},\mathrm{c}}^{\mathrm{s}}=1$ and $\mathcal{O} _{\mathrm{d},\mathrm{c}}^{\mathrm{s}}=\log _2\left( \frac{\zeta \rho }{4\pi}\sum_{y\in \mathcal{Y}_{\mathrm{c}}}{\sum_{z\in \mathcal{Z}_{\mathrm{c}}}{\delta _{\mathrm{c}}\left( y,z \right)}} \right)$.
\end{remark}

{Next} we consider the case when $N_y,N_z\rightarrow \infty$ and derive the following corollary.
\begin{corollary} \label{SC_CR_M_cor}
When $N_y,N_z\rightarrow \infty $, the asymptotic downlink CR under the S-C design is given by
\begin{align}\label{SC_CR_M}
\lim_{N_y,N_z\rightarrow \infty} \mathcal{R} _{\mathrm{d},\mathrm{c}}^{\mathrm{s}}=\log _2\left( 1+\frac{\mathcal{C}_{\rho}p\zeta}{3} \right)  .
\end{align}
\end{corollary}
\begin{IEEEproof}
The proof of this result is similar to that of \textbf{Corollary}~\ref{CC_CR_M_cor}.
\end{IEEEproof}

\begin{remark}
The results in \textbf{Corollary}~\ref{SC_CR_M_cor} indicates that as $N_y,N_z\rightarrow \infty $, {rather than growing unboundedly, the downlink CR under the S-C design converges to a finite quantity that is increasing in the AOR.}
\end{remark}

The downlink CR of the S-C design in the absence of polarization loss is studied in the following corollary.
\begin{corollary}\label{rsc_polar_cor}
When the polarization mismatch is eliminated, the downlink CR of the S-C design is given by 
\begin{align}\label{rsc_polar}
\bar{\mathcal{R}} _{\mathrm{d},\mathrm{c}}^{\mathrm{s}}\!=\!\log _2\!\Bigg(\!1\!+\!\frac{p\zeta\bar{\rho}}{4\pi}\sum_{y\in \mathcal{Y} _{\mathrm{c}}}{\sum_{z\in \mathcal{Z} _{\mathrm{c}}}\!{\arctan \!\bigg( \frac{yz}{\Psi _{\mathrm{c}}\sqrt{\Psi _{\mathrm{c}}^{2}\!+\!y^2\!+\!z^2}} \bigg)}} \!\Bigg).
\end{align}
When $N_y,N_z\rightarrow \infty $, we have
\begin{align}
\lim_{N_y,N_z\rightarrow \infty}\bar{\mathcal{R}} _{\mathrm{d},\mathrm{c}}^{\mathrm{s}}=\log _2\left( 1+\frac{\bar{\mathcal{C}}_\rho p\zeta}{2 } \right).  
\end{align}
\end{corollary}
\begin{IEEEproof}
The proof of this result is similar to that of \textbf{Corollary~\ref{rcc_polar_cor}}.
\end{IEEEproof}

Based on the analysis above for the downlink scenario, we can conclude following remarks. 
\vspace{-5pt}
\begin{remark}\label{do_compare}
In downlink near-field ISAC, we have $\mathcal{S} _{\mathrm{d},\mathrm{c}}^{\mathrm{c}}=\mathcal{S} _{\mathrm{d},\mathrm{c}}^{\mathrm{s}}$ and $\mathcal{S} _{\mathrm{d},\mathrm{s}}^{\mathrm{c}}=\mathcal{S} _{\mathrm{d},\mathrm{s}}^{\mathrm{s}}$, which indicates that the beamforming design does not impact the high-SNR slopes of both CR and SR. On the other hand, we have $\mathcal{O}_{\mathrm{d},\mathrm{c}}^{\mathrm{s}}-\mathcal{O} _{\mathrm{d},\mathrm{c}}^{\mathrm{c}}=\mathcal{O} _{\mathrm{d},\mathrm{s}}^{\mathrm{c}}-\mathcal{O} _{\mathrm{d},\mathrm{s}}^{\mathrm{s}}=\log_2\rho_{\mathrm{sc}}\in \left( -\infty ,0 \right] $, which indicates that the beamforming design influences the CR and SR by altering their high-SNR power offsets. The performance gaps
between the C-C design and the S-C design for both S\&C are influenced by the channel correlation.
\end{remark}

\vspace{-5pt}
\begin{remark}\label{do_constant}
Under the near-field channel model we investigated, {all CRs and SRs tend to finite quantities when $N\rightarrow \infty$. By contrast, as will be shown by the numerical results, CRs and SRs under the TCMs exhibit unbounded growth with $N$.}
\end{remark}

\subsection{Pareto Optimal Design}
In practical scenarios, the beamforming vector $\mathbf{w}$ can be customized to fulfill diverse quality of service requirements, creating a tradeoff between communication and sensing performance. To evaluate this tradeoff, we analyze the Pareto boundary of the SR-CR region achieved by the downlink near-field ISAC. The Pareto boundary comprises SR-CR pairs where it becomes impossible to augment one rate without simultaneously reducing the other \cite{pareto}. Particular, any rate pair located on the Pareto boundary of the SR-CR region can be determined by solving the following optimization problem:
\begin{align}\label{Problem_CR_SR_Tradeoff}
\max_{\mathbf{w},\mathcal{R}}\mathcal{R},\,\,\mathrm{s}.\mathrm{t}.\, \mathcal{R} _{\mathrm{d},\mathrm{s}}\ge \sigma \mathcal{R} ,\,\mathcal{R} _{\mathrm{d},\mathrm{c}}\ge \left( 1\!-\!\sigma \right) \mathcal{R} ,\,\left\| \mathbf{w} \right\| ^2\!=\!1,
\end{align}
where $\sigma\in\left[0,1\right]$ is a particular rate-profile parameter. The entire Pareto
boundary is obtained by solving the above problem with $\sigma$ varying from $0$ to $1$. Despite of its non-convexity, we can achieve an optimal closed-form solution for problem \eqref{Problem_CR_SR_Tradeoff} as follows.

\begin{figure} [t!]
\centering
\includegraphics[height=0.2\textwidth]{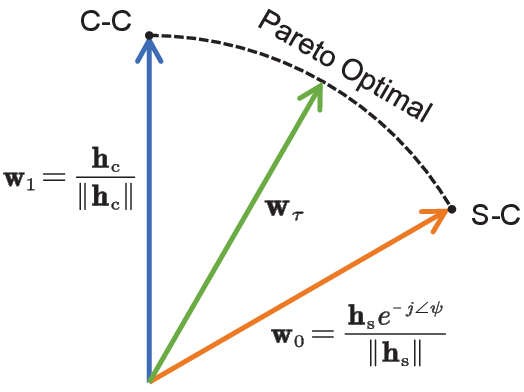}
 \caption{Pareto optimal beamforming vector.}
 \vspace{-5pt}
 \label{pareto}
\end{figure}

\begin{figure*}[!b]
\hrulefill
\setcounter{equation}{51}
\begin{align}\label{optimal_w}
\mathbf{w}_{\sigma}^{\star}=\!\begin{cases}
	\mathbf{w}_{\mathrm{c}},&		\sigma \in \left[ 0,\frac{\mathcal{R} _{\mathrm{d},\mathrm{s}}^{\mathrm{c}}}{\mathcal{R} _{\mathrm{d},\mathrm{c}}^{\mathrm{c}}+\mathcal{R} _{\mathrm{d},\mathrm{s}}^{\mathrm{c}}} \right]\\
	\frac{\mu _1\sqrt{\left( 2^{\left( 1-\sigma \right) \mathcal{R} ^{\star}}-1 \right) p}\mathbf{h}_{\mathrm{c}}+\mu _2\sqrt{\left( 2^{\sigma \!L\!\mathcal{R} ^{\star}}-1 \right) p}\xi \mathbf{h}_{\mathrm{s}}e^{-j\angle \left( \mathbf{h}_{\mathrm{c}}^{\mathsf{H}}\mathbf{h}_{\mathrm{s}} \right)}}{\left[ \left( 2^{( 1\!-\sigma ) \mathcal{R} ^{\star}}\!-1 \right) \mu _{1}^{2}p\left\| \mathbf{h}_{\mathrm{c}} \right\| ^2+\left( 2^{\sigma\! L\!\mathcal{R} ^{\star}}\!-1 \right) \mu _{2}^{2}p\xi ^2\left\| \mathbf{h}_{\mathrm{s}} \right\| ^2+2\sqrt{\left( 2^{( 1\!-\sigma) \mathcal{R}^{\star}}\!-1 \right) \left( 2^{\sigma L\mathcal{R} ^{\star}}\!-1 \right)}\mu _1\mu _2p\xi \left| \mathbf{h}_{\mathrm{c}}^{\mathsf{H}}\mathbf{h}_{\mathrm{s}} \right| \right] ^{\frac{1}{2}}},&		\sigma \in \left( \frac{\mathcal{R} _{\mathrm{d},\mathrm{s}}^{\mathrm{c}}}{\mathcal{R} _{\mathrm{d},\mathrm{c}}^{\mathrm{c}}+\mathcal{R} _{\mathrm{d},\mathrm{s}}^{\mathrm{c}}},\frac{\mathcal{R} _{\mathrm{d},\mathrm{s}}^{\mathrm{s}}}{\mathcal{R} _{\mathrm{d},\mathrm{c}}^{\mathrm{s}}+\mathcal{R} _{\mathrm{d},\mathrm{s}}^{\mathrm{s}}} \right)\\
	\mathbf{w}_{\mathrm{s}},&		\sigma \in \left[ \frac{\mathcal{R} _{\mathrm{d},\mathrm{s}}^{\mathrm{s}}}{\mathcal{R} _{\mathrm{d},\mathrm{c}}^{\mathrm{s}}+\mathcal{R} _{\mathrm{d},\mathrm{s}}^{\mathrm{s}}},1 \right]\\
\end{cases}.
\end{align}
\begin{equation}\label{equation_R}
\left( 2^{\sigma L\mathcal{R}}\!-1 \right) \left\| \mathbf{h}_{\mathrm{c}} \right\| ^2+\left( 2^{\left( 1-\sigma \right) \mathcal{R}}\!-1 \right) \xi ^2\left\| \mathbf{h}_{\mathrm{s}} \right\| ^2-2\sqrt{\left( 2^{\left( 1-\sigma \right) \mathcal{R}}\!-1 \right) \left( 2^{\sigma L\mathcal{R}}\!-1 \right)}\xi \left| \mathbf{h}_{\mathrm{c}}^{\mathsf{H}}\mathbf{h}_{\mathrm{s}} \right|=p\xi ^2\left( \left\| \mathbf{h}_{\mathrm{c}} \right\| ^2\left\| \mathbf{h}_{\mathrm{s}} \right\| ^2\!-\left| \mathbf{h}_{\mathrm{c}}^{\mathsf{H}}\mathbf{h}_{\mathrm{s}} \right|^2 \right) .
\end{equation}
\begin{subequations}
\begin{align}
\mu _1&=\frac{\xi ^2\left\| \mathbf{h}_{\mathrm{s}} \right\| ^2-\chi \xi \left| \mathbf{h}_{\mathrm{c}}^{\mathsf{H}}\mathbf{h}_{\mathrm{s}} \right|}{\left( \left\| \mathbf{h}_{\mathrm{c}} \right\| ^2-\chi ^{-1}\xi \left| \mathbf{h}_{\mathrm{c}}^{\mathsf{H}}\mathbf{h}_{\mathrm{s}} \right| \right) 2^{\sigma L\mathcal{R}^{\star}}\sigma L\ln 2+\left( \xi \left\| \mathbf{h}_{\mathrm{s}} \right\| ^2-\chi \left| \mathbf{h}_{\mathrm{c}}^{\mathsf{H}}\mathbf{h}_{\mathrm{s}} \right| \right) \xi 2^{\left( 1-\sigma \right) \mathcal{R}^{\star}}\left( 1-\sigma \right) \ln 2},\label{mu1}\\  
\mu _2&=\frac{\left\| \mathbf{h}_{\mathrm{c}} \right\| ^2-\chi ^{-1}\xi \left| \mathbf{h}_{\mathrm{c}}^{\mathsf{H}}\mathbf{h}_{\mathrm{s}} \right|}{\left( \left\| \mathbf{h}_{\mathrm{c}} \right\| ^2-\chi ^{-1}\xi \left| \mathbf{h}_{\mathrm{c}}^{\mathsf{H}}\mathbf{h}_{\mathrm{s}} \right| \right) 2^{\sigma L\mathcal{R}^{\star}}\sigma L\ln 2+\left( \xi \left\| \mathbf{h}_{\mathrm{s}} \right\| ^2-\chi \left| \mathbf{h}_{\mathrm{c}}^{\mathsf{H}}\mathbf{h}_{\mathrm{s}} \right| \right) \xi 2^{\left( 1-\sigma \right) \mathcal{R} ^{\star}}\left( 1-\sigma \right) \ln 2}.\label{mu2}
\end{align}
\end{subequations}
\end{figure*}

\begin{theorem}\label{pareto_theo}
For a given $\sigma$, by defining the auxiliary parameter $\xi\triangleq\sqrt{\frac{L\alpha _{\mathrm{s}}\zeta}{4\pi}\sum_{y\in \mathcal{Y} _{\mathrm{s}}}{\sum_{z\in \mathcal{Z} _{\mathrm{s}}}{\delta _{\mathrm{s}}\left( y,z \right)}}}$, the optimal beamforming vector is given by \eqref{optimal_w}, shown at the bottom of this page. Specifically, $\mathcal{R}^\star$ represents the optimal solution for $\mathcal{R}$, which is obtained by solving the equation \eqref{equation_R}. Moreover, $\mu_1$ and $\mu_2$ are Lagrange multipliers for problem \eqref{Problem_CR_SR_Tradeoff}, which are given in \eqref{mu1} and \eqref{mu2}, respectively, with $\chi =\sqrt{\frac{2^{\sigma L\mathcal{R} ^{\star}}-1}{2^{\left( 1-\sigma \right) \mathcal{R} ^{\star}}-1}}$. 
\end{theorem}
\begin{IEEEproof}
Please refer to Appendix~\ref{Appendix:C}.
\end{IEEEproof}

Based on the above results of the optimal beamforming vector, we can further deduce the following corollary.

\begin{corollary}\label{pareto_cor}
The Pareto boundary of the rate region can be attained through the beamforming vector as outlined below:
\setcounter{equation}{54}
\begin{align}
\mathbf{w}_{\tau}=\frac{\tau \mathbf{h}_{\mathrm{c}} +\left( 1-\tau \right) \mathbf{h}_{\mathrm{s}} {\rm{e}}^{-{\rm{j}}\angle \psi}}{\left\| \tau \mathbf{h}_{\mathrm{c}} +\left( 1-\tau \right) \mathbf{h}_{\mathrm{s}} {\rm{e}}^{-{\rm{j}}\angle \psi} \right\|},
\end{align}
where $\tau \in \left[ 0,1 \right] $ is the weighting factor.
\end{corollary}
\vspace{-3pt}
\begin{IEEEproof}
Please refer to Appendix~\ref{Appendix:D}.
\end{IEEEproof}

Noting that $\mathbf{w}_{\tau}$ can represent any arbitrary linear combination of $\mathbf{h}_{\mathrm{c}}$ and $\mathbf{h}_{\mathrm{s}} {\rm{e}}^{-{\rm{j}}\angle \psi}$ with non-negative real coefficients, we can draw the following conclusion.
\vspace{-5pt}
\begin{remark}\label{pareto_rem}
 The results in \textbf{Corollary}~\ref{pareto_cor} indicate that the Pareto optimal beamforming vector lies in the plane spanned by $\mathbf{h}_{\mathrm{c}}$ and $\mathbf{h}_{\mathrm{s}}{\rm{e}}^{-{\rm{j}}\angle \psi}$, as illustrated in Fig.~\ref{pareto}.
\end{remark}

Therefore, for a given value of $\tau$, let $\mathcal{R} _{\mathrm{d},\mathrm{s}}^{\tau}$ and $\mathcal{R} _{\mathrm{d},\mathrm{c}}^\tau$ represent the SR and CR on the Pareto boundary achieved by the corresponding optimal beamforming vector $\mathbf{w}_{\tau}$, respectively, which are given by 
\setcounter{equation}{55}
\begin{align}
&\mathcal{R} _{\mathrm{d},\mathrm{s}}^{\tau}=\frac{1}{L}\log _2\left( 1+pL\alpha _{\mathrm{s}}\left\| \mathbf{h}_{\mathrm{s}} \right\| ^2\times\right.\\
&\left.\frac{\tau ^2\rho \left\| \mathbf{h}_{\mathrm{c}} \right\| ^2\!\left\| \mathbf{h}_{\mathrm{s}} \right\| ^2\!+\!\left( 1\!-\!\tau \right) ^2\!\left\| \mathbf{h}_{\mathrm{s}} \right\| ^4\!+\!2\tau \!\left( 1\!-\!\tau \right) \left\| \mathbf{h}_{\mathrm{c}} \right\| \left\| \mathbf{h}_{\mathrm{s}} \right\| ^3}{\tau ^2\left\| \mathbf{h}_{\mathrm{c}} \right\| ^2+\left( 1-\tau \right) ^2\left\| \mathbf{h}_{\mathrm{s}} \right\| ^2+2\tau \left( 1-\tau \right) \left\| \mathbf{h}_{\mathrm{c}} \right\| \left\| \mathbf{h}_{\mathrm{s}} \right\|} \right)\! ,\notag\\
&\mathcal{R} _{\mathrm{d},\mathrm{c}}^{\tau}=\log _2\left( 1+p\times\right.\\
&\left.\frac{\tau ^2\!\left\| \mathbf{h}_{\mathrm{c}} \right\| ^4\!+\!\left( 1\!-\!\tau \right) ^2\!\rho \left\| \mathbf{h}_{\mathrm{c}} \right\| ^2\!\left\| \mathbf{h}_{\mathrm{s}} \right\| ^2\!+\!2\tau \!\left( 1\!-\!\tau \right) \left\| \mathbf{h}_{\mathrm{c}} \right\| ^3\!\left\| \mathbf{h}_{\mathrm{s}} \right\|}{\tau ^2\left\| \mathbf{h}_{\mathrm{c}} \right\| ^2+\left( 1-\tau \right) ^2\left\| \mathbf{h}_{\mathrm{s}} \right\| ^2+2\tau \left( 1-\tau \right) \left\| \mathbf{h}_{\mathrm{c}} \right\| \left\| \mathbf{h}_{\mathrm{s}} \right\|} \right)\! .\notag
\end{align}
In particular, we have $\left( \mathcal{R} _{\mathrm{d},\mathrm{s}}^{0},\mathcal{R} _{\mathrm{d},\mathrm{c}}^{0} \right) =\left( \mathcal{R} _{\mathrm{d},\mathrm{s}}^{\mathrm{s}},\mathcal{R} _{\mathrm{d},\mathrm{c}}^{\mathrm{s}} \right) $ and $\left( \mathcal{R} _{\mathrm{d},\mathrm{s}}^{1},\mathcal{R} _{\mathrm{d},\mathrm{c}}^{1} \right) =\left( \mathcal{R} _{\mathrm{d},\mathrm{s}}^{\mathrm{c}},\mathcal{R} _{\mathrm{d},\mathrm{c}}^{\mathrm{c}} \right) $. The closed-form expressions of $\mathcal{R} _{\mathrm{d},\mathrm{s}}^{\tau}$ and $\mathcal{R} _{\mathrm{d},\mathrm{c}}^\tau$ can be obtained by substituting the 
previously derived expressions for $\left\| \mathbf{h}_{\mathrm{c}} \right\| $ and $\left\| \mathbf{h}_{\mathrm{s}} \right\| $, which are omitted due to space limitations. Consequently, it can be easily shown that for any given $\tau$, the high-SNR slopes of SR and CR are the same as those achieved by the S-C and C-C design, i.e., any downlink SR-CR pair on the Pareto boundary exhibits identical high-SNR slopes. Additionally, all the SRs and CRs on the Pareto boundary converge to finite quantities when $N_y,N_z \rightarrow\infty$.

\subsection{Downlink Near-Field FDSAC}
We consider the near-field FDSAC as the baseline, where the bandwidth is split into two distinct sub-bands: one dedicated solely to sensing and the other designated for communication. Furthermore, the total power is also distributed into two separate portions, each allocated for the specific objectives of S\&C, respectively. In particular, we consider $\kappa\in \left[ 0,1 \right] $ fraction of the total bandwidth and $\iota\in \left[ 0,1 \right]  $ fraction of the total power is allocated to sensing. Accordingly, the downlink SR and CR of FDSAC are, respectively, given by
\begin{align}
&\mathcal{R} _{\mathrm{d},\mathrm{s}}^{\mathrm{f}}=\frac{\kappa}{L}\log _2\left( 1+\frac{\iota}{\kappa}pL\alpha _{\mathrm{s}}\left\| \mathbf{h}_{\mathrm{s}} \right\| ^4 \right) ,\\
&\mathcal{R} _{\mathrm{d},\mathrm{c}}^{\mathrm{f}}=\left( 1-\kappa \right) \log _2\left( 1+\frac{1-\iota}{1-\kappa}p\left\| \mathbf{h}_{\mathrm{c}} \right\| ^2 \right) .
\end{align}
It is worth noting that $(\mathcal{R} _{\mathrm{d},\mathrm{c}}^{\mathrm{f}},\mathcal{R} _{\mathrm{d},\mathrm{s}}^{\mathrm{f}})$ can be discussed in the way we discuss $(\mathcal{R} _{\mathrm{d},\mathrm{s}}^{\mathrm{s}},\mathcal{R}_{\mathrm{d},\mathrm{c}}^{\mathrm{c}})$. Upon concluding all the analyses of near-field ISAC and FDSAC, we consolidate the results of high-SNR slopes within Table~\ref{table1}.

\begin{remark} \label{FDSAC_compare}
The results in Table~\ref{table1} indicate that in the downlink near-field scenario, ISAC exhibits higher high-SNR slopes than FDSAC in terms of both CR and SR, which implies that ISAC offers greater degrees of freedom than FDSAC concerning both S\&C.  
\end{remark}
\vspace{-5pt}
\begin{table}[!t]
\center
\begin{tabular}{|c|c|c|}\hline
 System  & CR & SR \\ \hline
 ISAC (C-C) & $1$ &  $1/L$  \\ \hline
  ISAC (S-C) & $1$ & $1/L$  \\ \hline
 ISAC (Pareto Optimal) & $1$ &$1/L$  \\ \hline
 FDSAC & $\kappa$ & $\left(1-\kappa\right)/L$  \\ \hline
\end{tabular}
\caption{Downlink High-SNR Slopes}
\vspace{-5pt}
\label{table1}
\end{table}

\subsection{Rate Region Characterization}
The achievable downlink SR-CR region of ISAC and FDSAC systems are, respectively, given by
\begin{align}
\mathcal{C}_{\mathrm{d},\mathrm{i}}&=\!\left\{\left({\mathcal{R}}_{\mathrm{s}},{\mathcal{R}}_{\mathrm{c}}\right)|{\mathcal{R}}_{\mathrm{s}}\!\in\!\left[0,\mathcal{R}_{\mathrm{d},\mathrm{s}}^{\tau}\right],
{\mathcal{R}}_{\mathrm{c}}\!\in\!\left[0,\mathcal{R}_{\mathrm{d},\mathrm{c}}^{\tau}\right],\tau\!\in\!\left[0,1\right]\right\}\!,\label{Rate_Regio_ISAC}\\
\mathcal{C} _{\mathrm{d},\mathrm{f}}&=\!\left\{ \left( \mathcal{R} _{\mathrm{s}},\mathcal{R} _{\mathrm{c}} \right) \left|\!\! \begin{array}{c}
	\mathcal{R} _{\mathrm{s}}\in \!\left[ 0,\mathcal{R} _{\mathrm{d},\mathrm{s}}^{\mathrm{f}} \right] ,\mathcal{R} _{\mathrm{c}}\in \!\left[ 0,\mathcal{R} _{\mathrm{d},\mathrm{c}}^{\mathrm{f}} \right] ,\\
	\kappa \in \left[ 0,1 \right] ,\iota \in \left[ 0,1 \right]\\
\end{array} \right. \!\!\!\right\}  .\label{Rate_Regio_FDSAC}
\end{align}    
\vspace{-5pt}
\begin{theorem}\label{rate_region}
The regions described above satisfy $\mathcal{C} _{\mathrm{d},\mathrm{f}}\subseteq \mathcal{C} _{\mathrm{d},\mathrm{i}}$. 
\end{theorem}
\vspace{-3pt}
\begin{IEEEproof}
Please refer to Appendix~\ref{Proof_rate_region}.
\end{IEEEproof}

As per \textbf{Theorem}~\ref{rate_region}, the rate region attained by the donwlink near-field ISAC completely encompasses the region achieved by FDSAC. This can be primarily attributed to ISAC's integrated utilization of both spectrum and power resources.

\vspace{-3pt}
\section{Uplink Near-field ISAC} \label{uplink}
In this section, we investigate the performance of uplink near-field ISAC under the C-C and S-C designs according to the interference cancellation order of the SIC process. Also, the uplink achievable rate region is characterized with time-sharing strategy. The analysis for the scenario without polarization loss in the uplink case is similar to that of the downlink, which is omitted here for brevity, while the simulation results are presented in Section~\ref{numerical}.

\subsection{Communications-Centric Design}
In the context of the C-C design, the initial step involves estimating the target response signal by considering the communication signal as interference. Subsequently, the communication signals transmitted from the CU are detected after eliminating the influence of the sensing signal.

\subsubsection{Performance of Sensing}
From a worst-case design perspective, the aggregate interference-plus-noise $\mathbf{Z}_{\mathrm{c}}=\sqrt{p_{\mathrm{c}}}\mathbf{h}_{\mathrm{c}} \mathbf{s}_{\mathrm{c}}^{\mathsf{H}}+\mathbf{N}_{\mathrm{u}}$ is regarded as the Gaussian noise \cite{GaussianNoise}. In this case, the achievable SR is derived in the following theorem.

\begin{theorem}\label{up_CC_SR_the}
The SR of the uplink C-C design is given by 
\begin{align}
\mathcal{R} _{\mathrm{c},\mathrm{s}}^{\mathrm{c}}=\frac{1}{L}\log _2\!\left[ 1+p_{\mathrm{s}}L\alpha _{\mathrm{s}}\left\| \mathbf{h}_{\mathrm{s}} \right\| ^2\!\left( \left\| \mathbf{h}_{\mathrm{s}} \right\| ^2-\frac{p_{\mathrm{c}}\left| \mathbf{h}_{\mathrm{s}}^{\mathsf{H}}\mathbf{h}_{\mathrm{c}} \right|^2}{1\!+\!p_{\mathrm{c}}\!\left\| \mathbf{h}_{\mathrm{c}} \right\| ^2} \right) \right] .
\end{align}
\end{theorem}
\begin{IEEEproof}
Please refer to Appendix~\ref{Appendix:F}.
\end{IEEEproof}

It is worth to note that in most cases, the CU and target are located at different locations, leading to $\rho \ll 1$ in the near-field region \cite{channel_correlation}. Accordingly, we have
$\frac{p_{\mathrm{c}}\left| \mathbf{h}_{\mathrm{s}}^{\mathsf{H}}\mathbf{h}_{\mathrm{c}} \right|^2}{1+p_{\mathrm{c}}\left\| \mathbf{h}_{\mathrm{c}} \right\| ^2}<\frac{p_{\mathrm{c}}\left| \mathbf{h}_{\mathrm{s}}^{\mathsf{H}}\mathbf{h}_{\mathrm{c}} \right|^2}{p_{\mathrm{c}}\left\| \mathbf{h}_{\mathrm{c}} \right\| ^2}=\rho \left\| \mathbf{h}_{\mathrm{s}} \right\| ^2\ll \left\| \mathbf{h}_{\mathrm{s}} \right\| ^2$. As will be shown in Section \ref{Section_Uplink_S_C}, the SR achieved by the uplink S-C design is $\mathcal{R} _{\mathrm{c},\mathrm{s}}^{\mathrm{s}}=\frac{1}{L}\log _2\left( 1+p_{\mathrm{s}}L\alpha _{\mathrm{s}}\left\| \mathbf{h}_{\mathrm{s}} \right\| ^4 \right)$. The above facts suggest that the gap of SR between the uplink C-C and S-C design is negligible, i.e. $\mathcal{R} _{\mathrm{c},\mathrm{s}}^{\mathrm{c}}\approx \mathcal{R} _{\mathrm{c},\mathrm{s}}^{\mathrm{s}}$. This also implies that the near-field effect can be harnessed to effectively mitigate IFI, improving uplink ISAC performance. Furthermore, to unveil the system's lower-bound performance, we consider the worst case where $\rho =1$, i.e., $\left| \mathbf{h}_{\mathrm{s}}^{\mathsf{H}}\mathbf{h}_{\mathrm{c}} \right|^2=\left\| \mathbf{h}_{\mathrm{c}} \right\| ^2\left\| \mathbf{h}_{\mathrm{s}} \right\| ^2$. In this case, the SR is written as
\begin{align}\label{R_cs_bound}
\Tilde{\mathcal{R}} _{\mathrm{c},\mathrm{s}}^{\mathrm{c}}=\frac{1}{L}\log _2\left( 1+\frac{p_{\mathrm{s}}L\alpha _{\mathrm{s}}\left\| \mathbf{h}_{\mathrm{s}} \right\| ^4}{1+p_{\mathrm{c}}\left\| \mathbf{h}_{\mathrm{c}} \right\| ^2} \right) .
\end{align}
The following corollary provides an exact closed-form expression of $\Tilde{\mathcal{R}} _{\mathrm{c},\mathrm{s}}^{\mathrm{c}}$ and its high-SNR approximation.

\begin{corollary}
The SR achieved by the uplink C-C design is lower bounded by 
\begin{align}
\Tilde{\mathcal{R}} _{\mathrm{c},\mathrm{s}}^{\mathrm{c}}\!=\!\frac{1}{L}\!\log _2\!\!\left(\! 1\!+\!\frac{p_{\mathrm{s}}L\alpha _{\mathrm{s}}\zeta ^2\left( \sum_{y\in \mathcal{Y} _{\mathrm{s}}}{\sum_{z\in \mathcal{Z} _{\mathrm{s}}}{\delta _{\mathrm{s}}\left( y,z \right)}} \right) ^2}{16\pi ^2\!+\!4\pi p_{\mathrm{c}}\zeta \sum_{y\in \mathcal{Y} _{\mathrm{c}}}\!{\sum_{z\in \mathcal{Z} _{\mathrm{c}}}\!{\delta _{\mathrm{c}}\!\left( y,z \right)}}} \!\right).    
\end{align}
For large $p$, its high-SNR approximation is given by
\begin{align}
&{\Tilde{\mathcal{R}} _{\mathrm{c},\mathrm{s}}^{\mathrm{c}}\approx}\frac{1}{L}\left[ \log _2p_{\mathrm{s}}~+\right.\notag\\
&~~~\left.\log _2\left( \frac{L\alpha _{\mathrm{s}}\zeta ^2\left( \sum_{y\in \mathcal{Y} _{\mathrm{s}}}{\sum_{z\in \mathcal{Z} _{\mathrm{s}}}{\delta _{\mathrm{s}}\left( y,z \right)}} \right) ^2}{16\pi ^2\!+\!4\pi p_{\mathrm{c}}\zeta \sum_{y\in \mathcal{Y} _{\mathrm{c}}}\!{\sum_{z\in \mathcal{Z} _{\mathrm{c}}}{\delta _{\mathrm{c}}\!\left( y,z \right)}}} \right) \right],
\end{align}
indicating a hign-SNR slope of $\frac{1}{L}$.
\end{corollary}
\vspace{-3pt}
\begin{IEEEproof}
The proof of this result is similar to that of \textbf{Theorem}~\ref{CC_CR_theorem}.
\end{IEEEproof}
Then, similar to the analysis of downlink scenario, we also investigate the asymptotic uplink performance for the case when $N_y,N_z\rightarrow\infty$.

\begin{corollary}
When the number of array elements goes to infinity, we have
\begin{align}
\underset{N_y,N_z\rightarrow \infty}{\lim}\Tilde{\mathcal{R}} _{\mathrm{c},\mathrm{s}}^{\mathrm{c}}=\frac{1}{L}\log _2\left( 1+\frac{p_{\mathrm{s}}L\alpha _{\mathrm{s}}\zeta ^2}{9+3p_{\mathrm{c}}\zeta} \right) ,
\end{align}
{which is a finite quantity.}
\end{corollary}
\begin{IEEEproof}
The proof of this result is similar to that of \textbf{Corollary}~\ref{CC_CR_M_cor}.
\end{IEEEproof}

\subsubsection{Performance of Communications}
After the estimation of the target response, the echo signal $\mathbf{GX}$ will be removed from the received superposed S\&C signal. The remained communication signal can be then directly detected with the optimal detection vector $\frac{\mathbf{h}^{*}_{\mathrm{c}}}{\left\| \mathbf{h}_{\mathrm{c}} \right\|}$ without interference, which yields a similar CR to the downlink C-C design by simply replacing $p_{\mathrm{s}}$ with $p_\mathrm{c}$, i.e., $\mathcal{R} _{\mathrm{c},\mathrm{c}}^{\mathrm{c}}=\log _2\left( 1+p_{\mathrm{c}}\left\| \mathbf{h}_{\mathrm{c}} \right\| ^2 \right)$.

\subsection{Sensing-Centric Design}
In the S-C design, the BS initially detects the communication signal, treating the echo signal $\mathbf{GX}$ as interference. Subsequently, the BS subtracts the detected communication signal from the received signal, using the remaining part for sensing purpose.

\subsubsection{Performance of Communications}
From a worst-case design perspective, the aggregate interference-plus-noise  $\mathbf{Z}_{\mathrm{s}}=\sqrt{p_{\mathrm{s}}}\mathbf{Gws}_{\mathrm{s}}^{\mathsf{H}}+\mathbf{N}_{\mathrm{u}}$ can be treated as the Gaussian noise \cite{GaussianNoise}. Based on this, the uplink CR of the near-field ISAC is given in the following theorem. 

\begin{theorem}\label{up_SC_CR_the}
In the S-C design, the uplink CR is given by
\begin{align}\label{up_SC_CR}
\mathcal{R} _{\mathrm{c},\mathrm{c}}^{\mathrm{s}}\!=\!\log _2\!\left[ 1\!+\!p_{\mathrm{c}}\left( \left\| \mathbf{h}_{\mathrm{c}} \right\| ^2\!-\!\frac{p_{\mathrm{s}}\alpha _{\mathrm{s}}\left\| \mathbf{h}_{\mathrm{s}} \right\| ^2\left| \mathbf{h}_{\mathrm{c}}^{\mathsf{H}}\mathbf{h}_{\mathrm{s}} \right|^2}{1+p_{\mathrm{s}}\alpha _{\mathrm{s}}\left\| \mathbf{h}_{\mathrm{s}} \right\| ^4} \right) \right] .  
\end{align}
\end{theorem}
\begin{IEEEproof}
Please refer to Appendix~\ref{Appendix:G}.
\end{IEEEproof}

Given that $\frac{p_{\mathrm{s}}\alpha _{\mathrm{s}}\left\| \mathbf{h}_{\mathrm{s}} \right\| ^2\left| \mathbf{h}_{\mathrm{c}}^{\mathsf{H}}\mathbf{h}_{\mathrm{s}} \right|^2}{1+p_{\mathrm{s}}\alpha _{\mathrm{s}}\left\| \mathbf{h}_{\mathrm{s}} \right\| ^4}<\frac{p_{\mathrm{s}}\alpha _{\mathrm{s}}\left\| \mathbf{h}_{\mathrm{s}} \right\| ^2\left| \mathbf{h}_{\mathrm{c}}^{\mathsf{H}}\mathbf{h}_{\mathrm{s}} \right|^2}{p_{\mathrm{s}}\alpha _{\mathrm{s}}\left\| \mathbf{h}_{\mathrm{s}} \right\| ^4}=\rho \left\| \mathbf{h}_{\mathrm{c}} \right\| ^2\ll \left\| \mathbf{h}_{\mathrm{c}} \right\| ^2$ in the near-field region, we have $\mathcal{R} _{\mathrm{c},\mathrm{c}}^{\mathrm{s}}\approx \mathcal{R} _{\mathrm{c},\mathrm{c}}^{\mathrm{c}}$. Furthermore, we provide a lower bound of the uplink CR under the S-C design for $\rho=1$, which is expressed as 
\begin{align}\label{R_sc_bound}
\Tilde{\mathcal{R}} _{\mathrm{c},\mathrm{c}}^{\mathrm{s}}=\log _2\left( 1+\frac{p_{\mathrm{c}}\left\| \mathbf{h}_{\mathrm{c}} \right\| ^2}{1+p_{\mathrm{s}}\alpha _{\mathrm{s}}\left\| \mathbf{h}_{\mathrm{s}} \right\| ^4} \right) .    
\end{align}
The following corollary provides an exact closed-form expression of $\Tilde{\mathcal{R}} _{\mathrm{c},\mathrm{c}}^{\mathrm{s}}$ and its high-SNR approximation.

\begin{corollary}
{The CR achieved by the uplink S-C design is lower bounded by}
\begin{align}
\Tilde{\mathcal{R}} _{\mathrm{c},\mathrm{c}}^{\mathrm{s}}\!=\!\log _2\!\!\left(\! 1\!+\!\frac{4\pi p_{\mathrm{c}}\zeta \sum_{y\in \mathcal{Y} _{\mathrm{c}}}{\sum_{z\in \mathcal{Z} _{\mathrm{c}}}{\delta _{\mathrm{c}}\left( y,z \right)}}}{16\pi ^2\!+\!p_{\mathrm{s}}\alpha _{\mathrm{s}}\zeta ^2\!\left( \sum_{y\in \mathcal{Y} _{\mathrm{s}}}\!{\sum_{z\in \mathcal{Z} _{\mathrm{s}}}\!{\delta _{\mathrm{s}}\left( y,z \right)}} \right) ^2}\! \right) .    
\end{align}
{For large $p_{\mathrm{c}}$}, its high-SNR approximation is given by
\begin{align}
&{\Tilde{\mathcal{R}} _{\mathrm{c},\mathrm{c}}^{\mathrm{s}}\approx}\log _2p_{\mathrm{c}}~+\notag\\
&~~~~\log _2\!\left(\frac{4\pi \zeta \sum_{y\in \mathcal{Y} _{\mathrm{c}}}{\sum_{z\in \mathcal{Z} _{\mathrm{c}}}{\delta _{\mathrm{c}}\left( y,z \right)}}}{16\pi ^2\!+\!p_{\mathrm{s}}\alpha _{\mathrm{s}}\zeta ^2\!\left( \sum_{y\in \mathcal{Y} _{\mathrm{s}}}\!{\sum_{z\in \mathcal{Z} _{\mathrm{s}}}\!{\delta _{\mathrm{s}}\left( y,z \right)}} \right) ^2}\right),
\end{align}
indicating a hign-SNR slope of one.
\end{corollary}
\vspace{-3pt}
\begin{IEEEproof}
The proof of this result is similar to that of \textbf{Theorem}~\ref{CC_CR_theorem}.
\end{IEEEproof}

\begin{corollary} \label{up_SC_CR_N_cor}
When $N_y,N_z\rightarrow \infty$, the asymptotic expression of $\Tilde{\mathcal{R}} _{\mathrm{c},\mathrm{c}}^{\mathrm{s}}$ follows
\begin{align}\label{up_SC_CR_N}
\underset{N_y,N_z\rightarrow \infty}{\lim}\Tilde{\mathcal{R}} _{\mathrm{c},\mathrm{c}}^{\mathrm{s}}=\log _2\left( 1+\frac{3p_{\mathrm{c}}\zeta}{9+p_{\mathrm{s}}\alpha _{\mathrm{s}}\zeta ^2} \right)  ,
\end{align}
{which is a finite quantity.}
\end{corollary}
\vspace{-3pt}
\begin{IEEEproof}
The proof of this result is similar to that of \textbf{Corollary}~\ref{CC_CR_M_cor}.
\end{IEEEproof}

\subsubsection{Performance of Sensing}\label{Section_Uplink_S_C}
After the decoded communication signal is removed, the rest part can be directly used for sensing without interference. It can be easily shown that under this circumstance, sensing yields the same performance as in the downlink S-C design, i.e., $\mathcal{R} _{\mathrm{c},\mathrm{s}}^{\mathrm{s}}=\frac{1}{L}\log _2\left( 1+p_{\mathrm{s}}L\alpha _{\mathrm{s}}\left\| \mathbf{h}_{\mathrm{s}} \right\| ^4 \right) $.

Based on the analysis of the near-field uplink ISAC above, we can draw the following conclusion.

\begin{remark} \label{up_compare}
The SIC order does not influence the high-SNR slopes for either CR or SR, but it has an effect on the S\&C performance by altering the high-SNR power offsets.
\end{remark}

\begin{table}[!t]
\center
\begin{tabular}{|c|c|c|}\hline
 System  & CR & SR \\ \hline
 ISAC (C-C) & $1$ &  $1/L$  \\ \hline
  ISAC (S-C) & $1$ & $1/L$  \\ \hline
 ISAC (Time-Sharing) & $1$ &$1/L$  \\ \hline
 FDSAC & $\kappa$ & $\left(1-\kappa\right)/L$  \\ \hline
\end{tabular}
\caption{Uplink High-SNR Slopes}
\vspace{-5pt}
\label{table2}
\end{table}

\subsection{Rate Region Characterization}
We now define the uplink SR-CR region achieved by the near-field ISAC. By employing the time-sharing strategy \cite{mimo}, we implement the S-C design with probability $\varrho $ and the C-C design with probability $1-\varrho $. For a given $\varrho$, the attainable rate pair is denoted as $\left( \mathcal{R} _{\mathrm{c},\mathrm{s}}^{\varrho },\mathcal{R} _{\mathrm{c},\mathrm{c}}^{\varrho } \right) $, where $\mathcal{R} _{\mathrm{c},\mathrm{s}}^{\varrho }=\varrho  \mathcal{R} _{\mathrm{c},\mathrm{s}}^{\mathrm{s}}+\left( 1-\varrho  \right) \mathcal{R} _{\mathrm{c},\mathrm{s}}^{\mathrm{c}}$ and $\mathcal{R} _{\mathrm{c},\mathrm{c}}^{\varrho }=\varrho  \mathcal{R} _{\mathrm{c},\mathrm{c}}^{\mathrm{s}}+\left( 1-\varrho  \right) \mathcal{R} _{\mathrm{c},\mathrm{c}}^{\mathrm{c}}$. Therefore, the achievable SR-CR region of uplink ISAC satisfies
\begin{align}
\mathcal{C} _{\mathrm{c},\mathrm{i}}\!=\!\left\{ \left( \mathcal{R} _{\mathrm{s}},\mathcal{R} _{\mathrm{c}} \right) \left| \mathcal{R} _{\mathrm{s}}\!\in \!\left[ 0,\mathcal{R} _{\mathrm{c},\mathrm{s}}^{\varrho} \right] ,\mathcal{R} _{\mathrm{c}}\!\in \!\left[ 0,\mathcal{R} _{\mathrm{c},\mathrm{c}}^{\varrho} \right] ,\varrho \in \left[ 0,1 \right] \right. \! \right\} .
\end{align}
\vspace{-5pt}
\begin{remark}
By exploiting the sandwich theorem, we can obtain that any rate pair achieved by the time-sharing strategy yields the same high-SNR slopes.
\end{remark}

\begin{figure*}[!t]
    \centering
    \subfigbottomskip=5pt
	\subfigcapskip=0pt
\setlength{\abovecaptionskip}{0pt}
    \subfigure[CR vs. $p$.]
    {
        \includegraphics[height=0.285\textwidth]{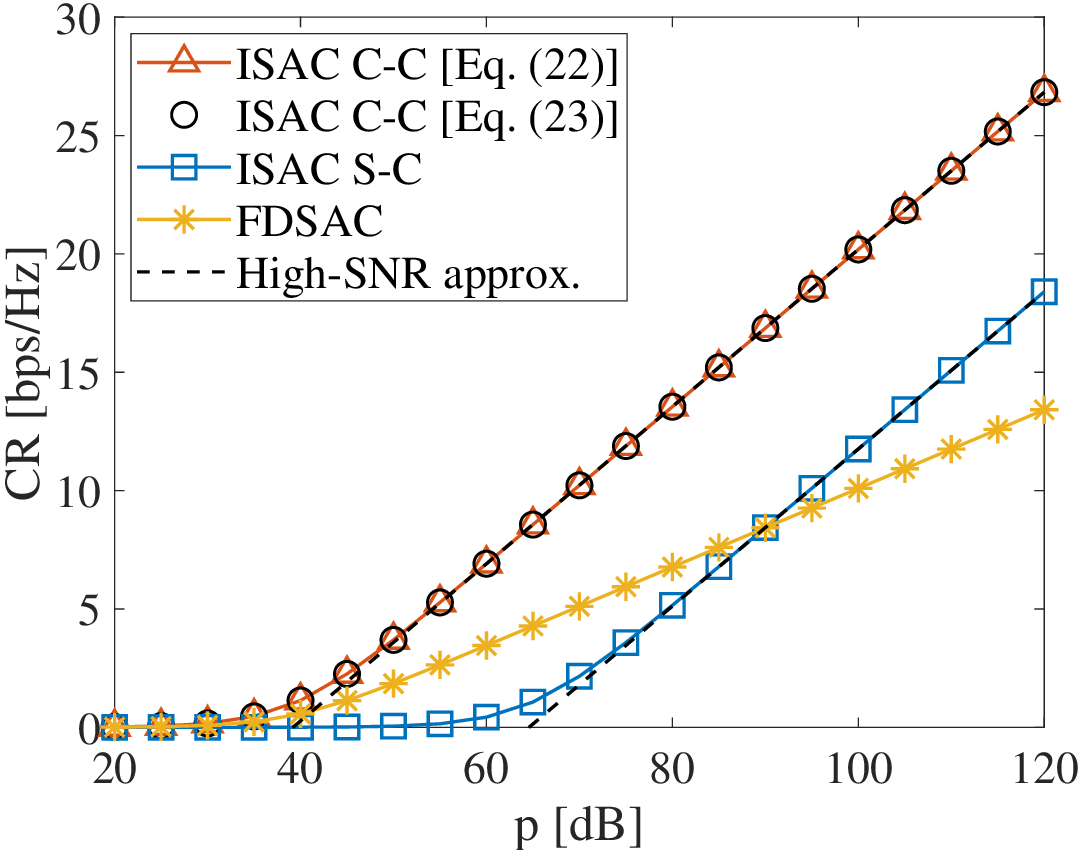}
	   \label{do_CRvsp}	
    }
    \quad
    \subfigure[SR vs. $p$.]
    {
        \includegraphics[height=0.285\textwidth]{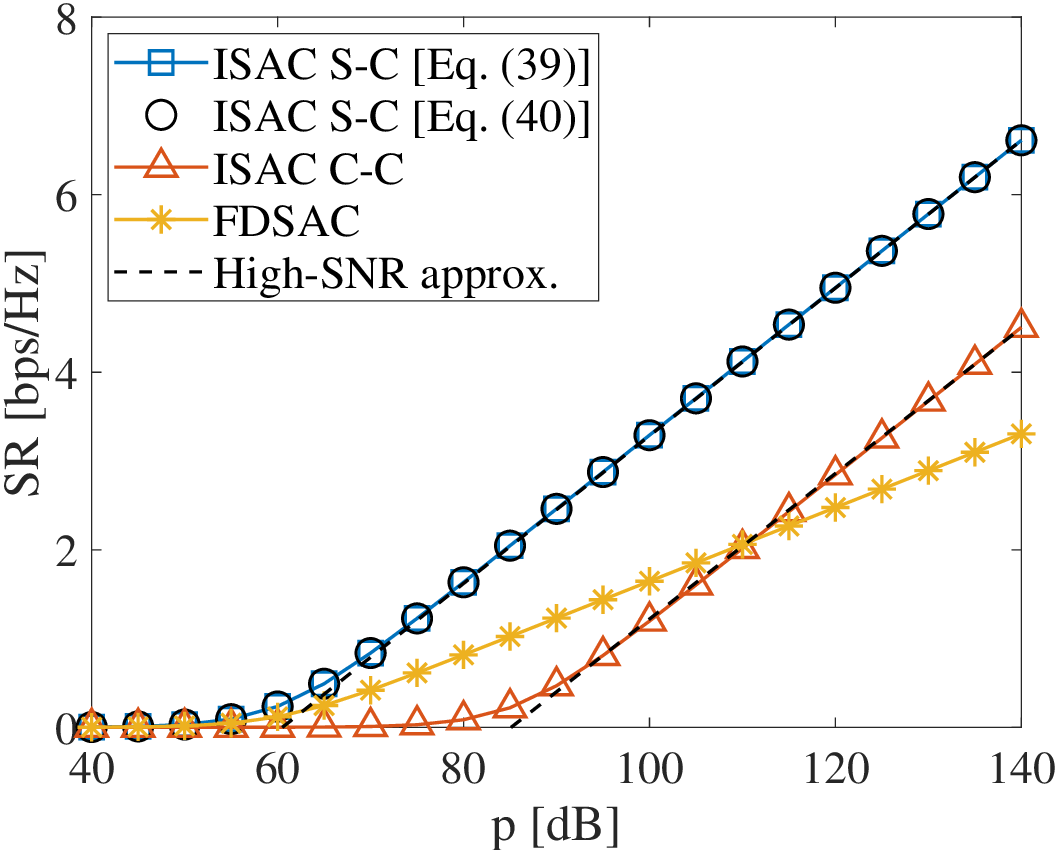}
	   \label{do_SRvsp}	
    }
\caption{Downlink performance versus SNR.}
    \label{do_SNR}
\end{figure*}

\begin{figure*}[!t]
    \centering
    \subfigbottomskip=5pt
	\subfigcapskip=0pt
\setlength{\abovecaptionskip}{0pt}
    \subfigure[CR vs. $N$.]
    {
        \includegraphics[height=0.28\textwidth]{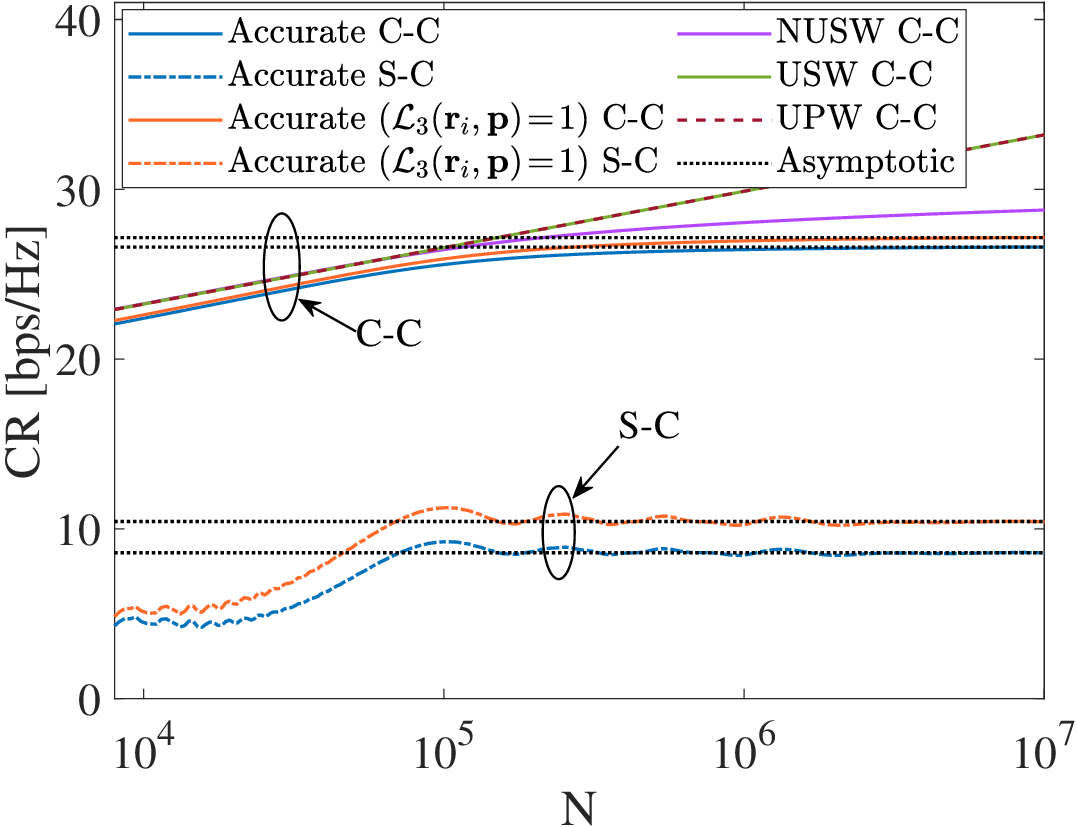}
	   \label{do_CRvsN}	
    }
    \quad
   \subfigure[SR vs. $N$.]
    {
        \includegraphics[height=0.285\textwidth]{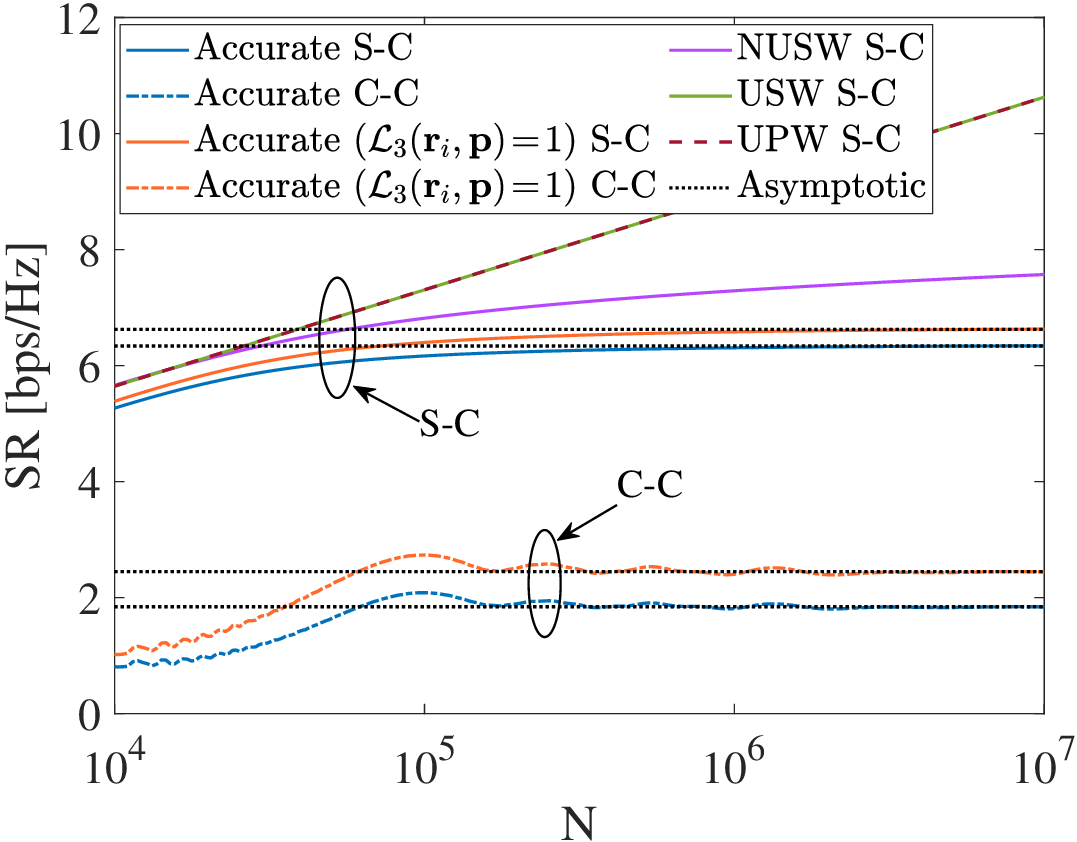}
	   \label{do_SRvsN}	
    }
\caption{Downlink performance versus number of array elements.}
    \label{do_N}
\end{figure*}

\subsection{Performance of FDSAC}
The uplink FDSAC is served as baseline, where a fraction $\kappa \in \left[0,1\right]$ of the total bandwidth is designated for sensing, while the remaining fraction is designated for communications. Accordingly, the SR and CR of FDSAC are given by
\begin{align}
&\mathcal{R} _{\mathrm{c},\mathrm{s}}^{\mathrm{f}}=\frac{\kappa}{L}\log _2\left( 1+\frac{p_{\mathrm{s}}L\alpha _{\mathrm{s}}\left\| \mathbf{h}_{\mathrm{s}} \right\| ^4}{\kappa}\right)  ,\\
&\mathcal{R} _{\mathrm{c},\mathrm{c}}^{\mathrm{f}}=\left( 1-\kappa \right) \log _2\left( 1+\frac{p_{\mathrm{c}}\left\| \mathbf{h}_{\mathrm{c}} \right\| ^2}{1-\kappa} \right)  .
\end{align}
It is worth noting that $(\mathcal{R} _{\mathrm{c},\mathrm{c}}^{\mathrm{f}},\mathcal{R} _{\mathrm{c},\mathrm{s}}^{\mathrm{f}})$ can be discussed in the way we discuss $(\mathcal{R} _{\mathrm{c},\mathrm{c}}^{\mathrm{c}},\mathcal{R} _{\mathrm{c},\mathrm{s}}^{\mathrm{s}})$. After completing all the analyses of the uplink case, we summarize the results pertaining to the high-SNR slope in Table~\ref{table2}.

\begin{remark}\label{up_FDSAC_compare}
The results in Table~\ref{table2} indicate that uplink ISAC achieves larger high-SNR slopes than FDSAC in terms of both SR and CR.
\end{remark}

\begin{figure*}[!t]
    \centering
    \subfigbottomskip=5pt
	\subfigcapskip=0pt
\setlength{\abovecaptionskip}{0pt}
    \subfigure[CR (C-C) vs. $r$.]
    {
        \includegraphics[height=0.28\textwidth]{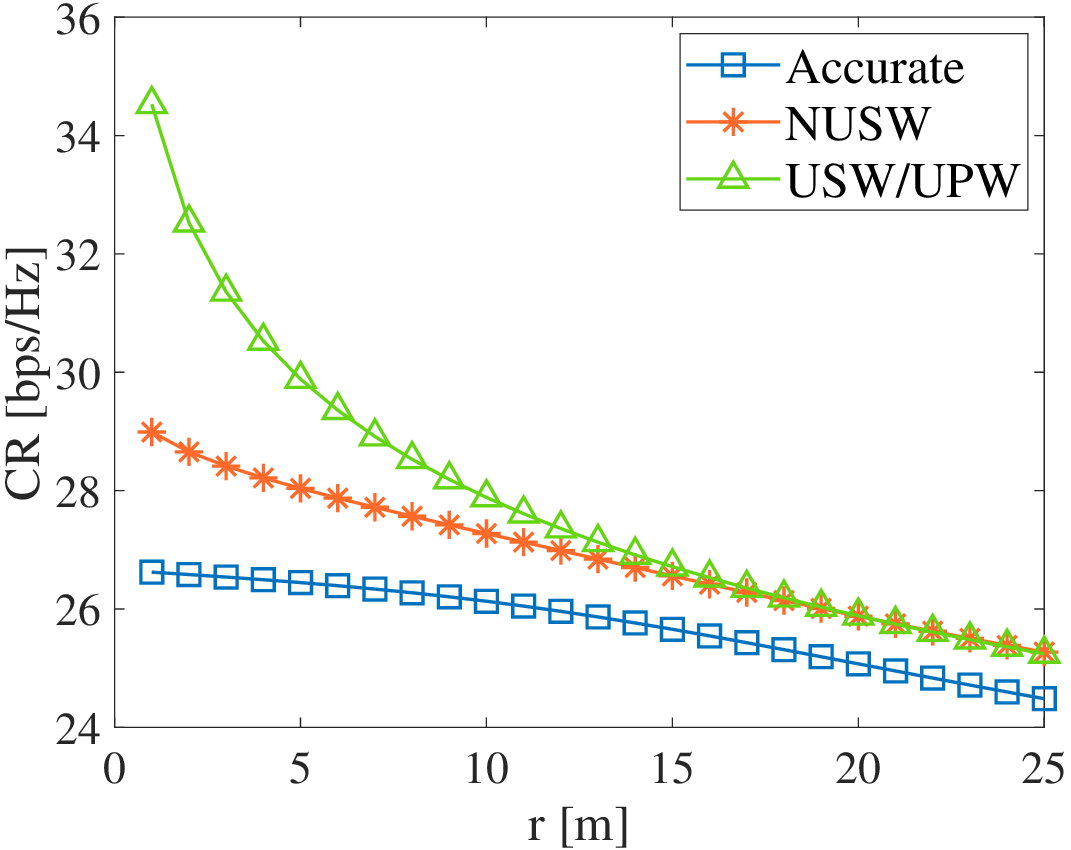}
	   \label{CRvsr}	
    }
   \quad
   \subfigure[SR (S-C) vs. $r$.]
    {
        \includegraphics[height=0.28\textwidth]{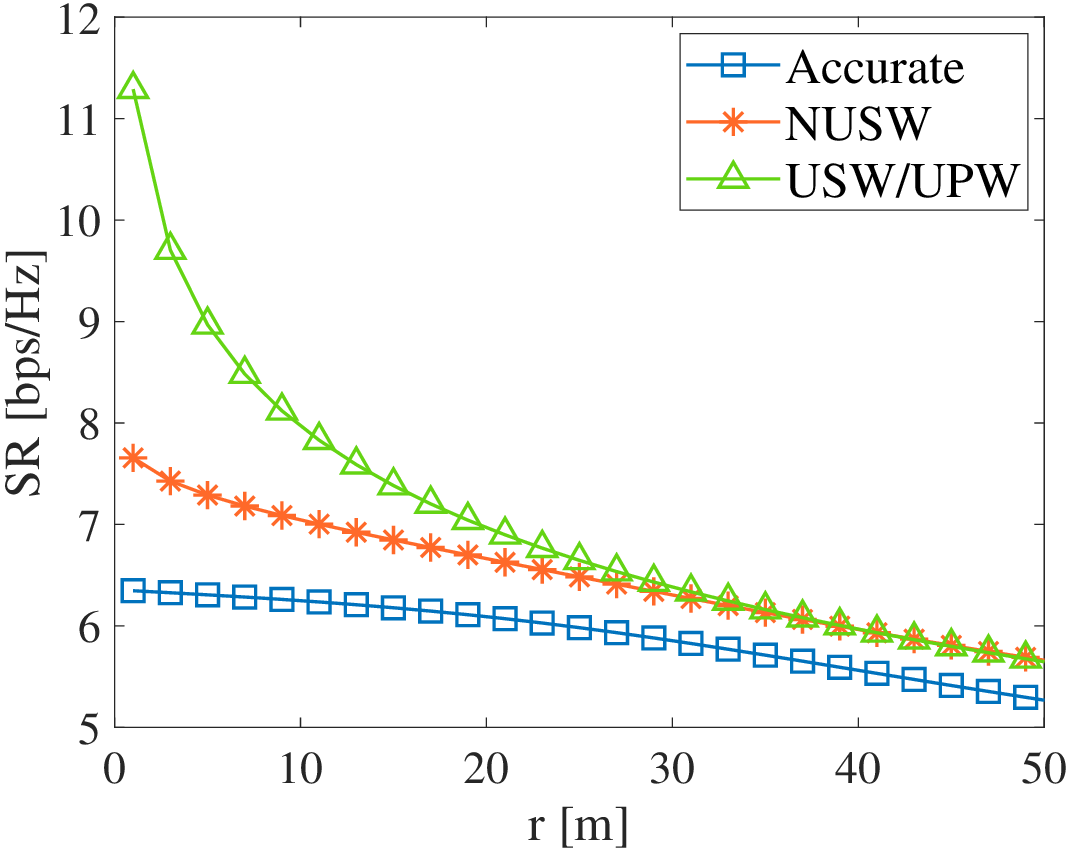}
	   \label{SRvsr}	
    }
\caption{Downlink performance versus distance with $N_y=N_z=1001$.}
\vspace{-2pt}
    \label{do_r}
\end{figure*}

\begin{figure} [!t]
\centering
\includegraphics[height=0.3\textwidth]{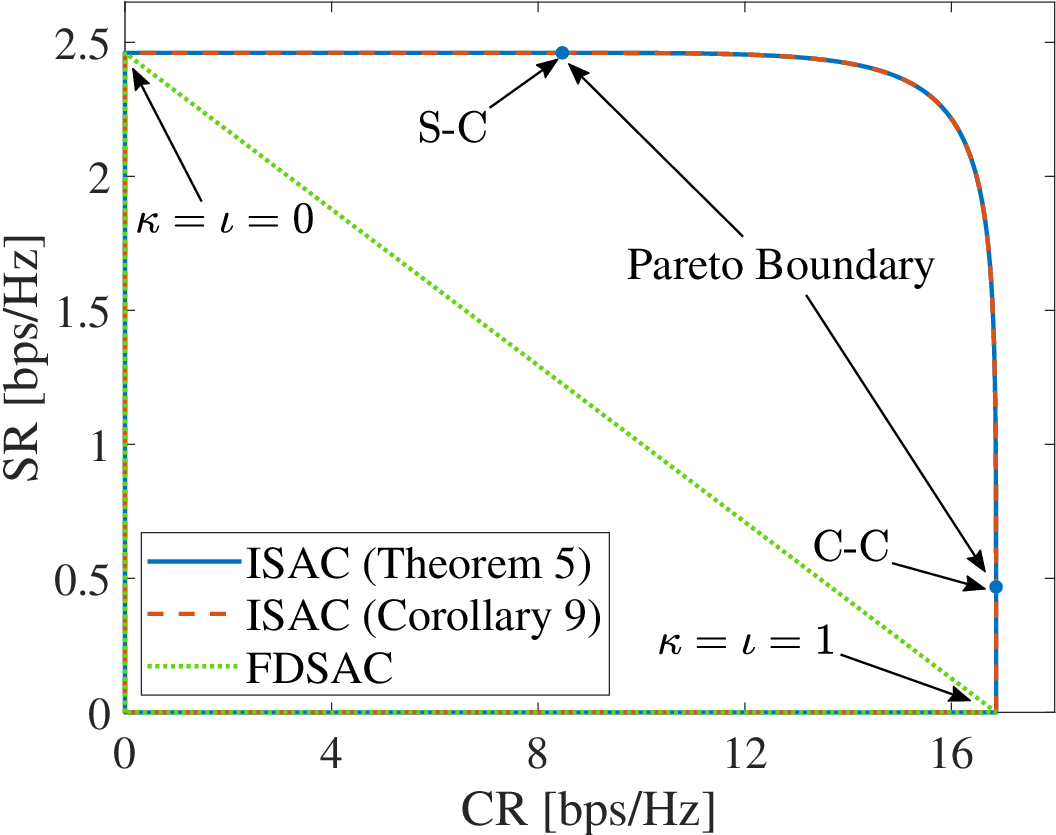}
 \caption{Downlink rate regions.}
 \vspace{-2pt}
 \label{do_region}
\end{figure}

\section{Numerical Results} \label{numerical}
In this section, numerical results for the S\&C performance of the near-field ISAC systems are presented. Without otherwise specification, the simulation parameter settings are defined as follows: $\lambda =0.125$ m, $d=\frac{\lambda}{2}$ m, $A=\frac{\lambda ^2}{4\pi}$, $L=4$, $\alpha_s=1$, $\kappa=\iota=0.5$, $N_y=N_z=15$, $p=90$ dB, $\left( r_{\mathrm{c}},\theta _{\mathrm{c}},\phi _{\mathrm{c}} \right) =\left( 10 \ \text{m},\frac{\pi}{4},\frac{\pi}{6} \right) $, and $\left( r_{\mathrm{s}},\theta _{\mathrm{s}},\phi _{\mathrm{s}} \right) =\left( 5 \ \text{m},\frac{\pi}{4}, -\frac{\pi}{6} \right) $.

\begin{figure*}[!t]
    \centering
    \subfigbottomskip=5pt
	\subfigcapskip=0pt
\setlength{\abovecaptionskip}{0pt}
    \subfigure[CR vs. $p_{\mathrm{c}}$ with $p_{\mathrm{s}}=85$ dB.]
    {
        \includegraphics[height=0.28\textwidth]{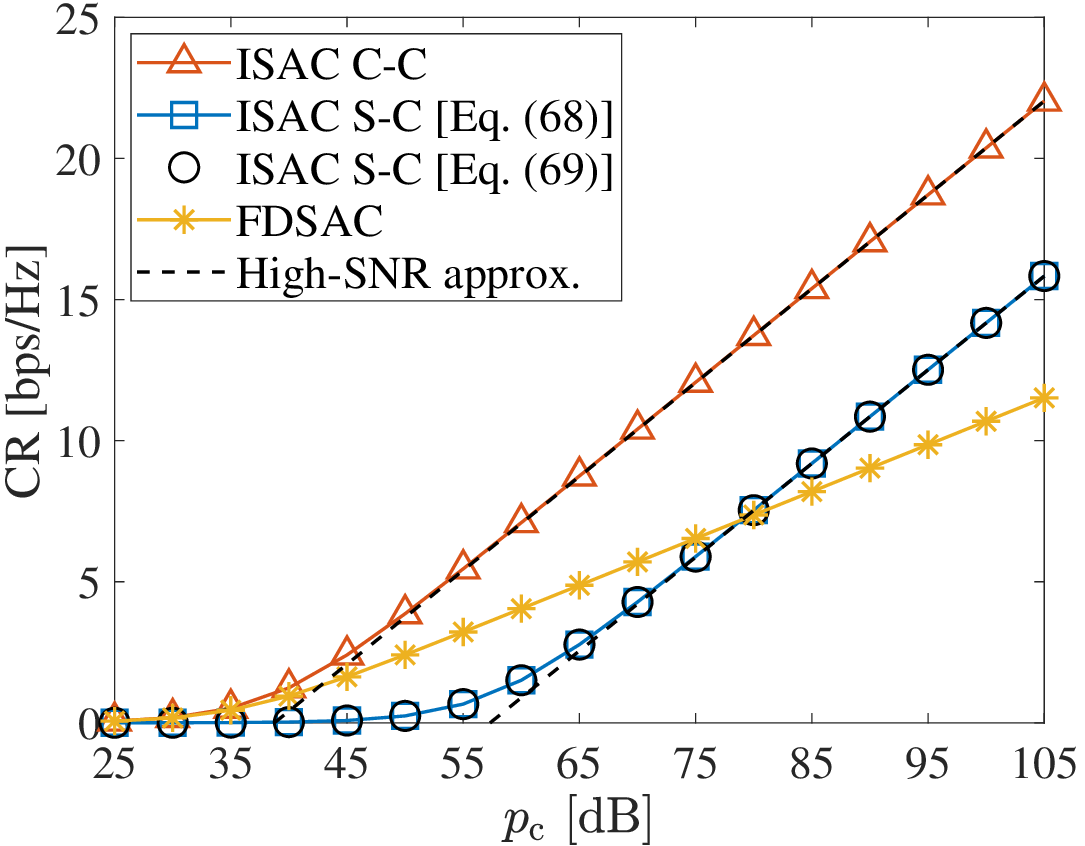}
	   \label{up_CRvsp}	
    }
   \quad
    \subfigure[SR vs. $p_{\mathrm{s}}$ with $p_{\mathrm{c}}=60$ dB]
    {
        \includegraphics[height=0.28\textwidth]{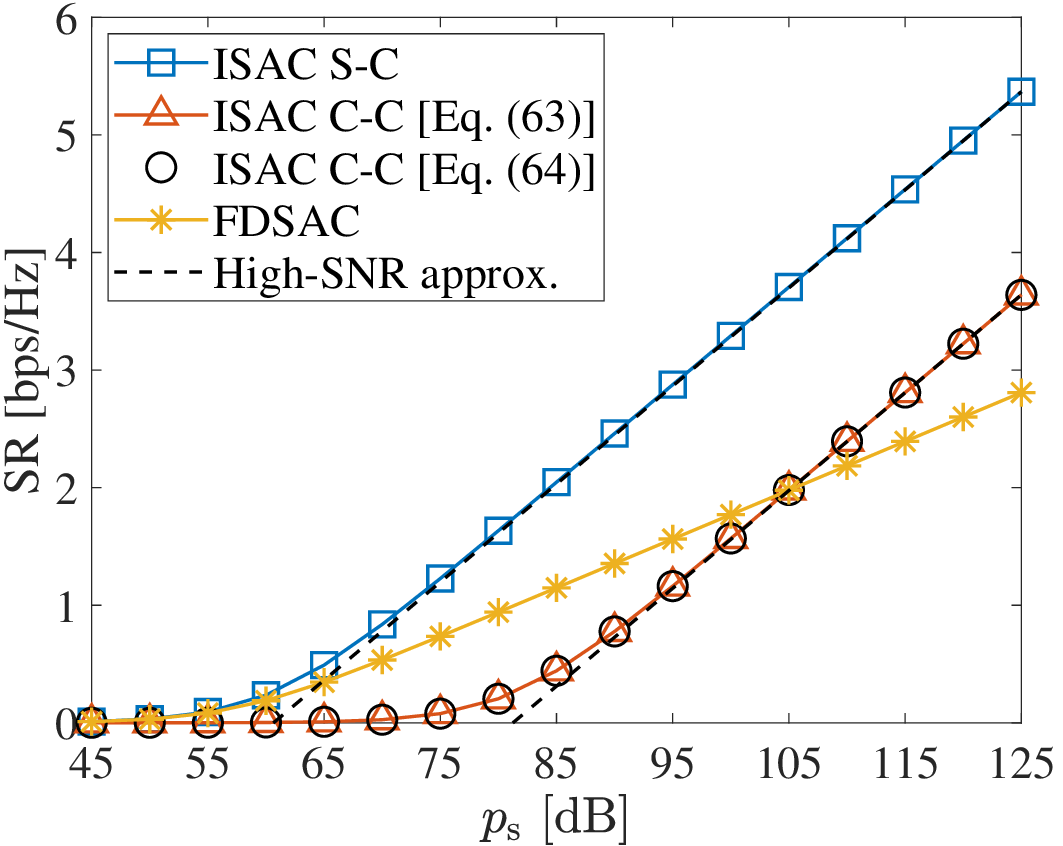}
	   \label{up_SRvsp}	
    }
\caption{Uplink performance versus SNR.}
    \label{up_SNR}
    \vspace{-2pt}
\end{figure*}

\subsection{Downlink}
Fig.~\ref{do_CRvsp} and Fig.~\ref{do_SRvsp} plot the downlink CRs and SRs versus the transmit SNR $p$, respectively. It is evident that C-C ISAC attains the best communication performance, while S-C ISAC records the best sensing performance. The derived closed-form results match the simulation results well, and the high-SNR approximations precisely track the results in the high-SNR region. We can also observe that the C-C ISAC and the S-C ISAC have the same high-SNR slopes in terms of both CR and SR, corroborating the statement in \textbf{Remark}~\ref{do_compare}. Importantly, both of these high-SNR slopes surpass those achieved by FDSAC.

Fig.~\ref{do_CRvsN} and Fig.~\ref{do_SRvsN} respectively illustrate the changes in CRs and SRs for various channel models as the number of UPA antennas varies. We can observe that as $N$ increases, the CRs achieved by our accurate model and the TCMs follow distinctly different scaling patterns. The CRs and SRs achieved by the accurate model converge to upper limits accurately tracked by our derived asymptotic results, which is aligned with \textbf{Remark}~\ref{do_constant}. In contrast, the CRs of the C-C design and the SRs of the S-C design achieved by the TCMs exhibit unbounded growth with $N$. This unrestrained increase is primarily attributed to the neglect of aperture loss, leading to the potential violation of the energy-conservation laws. Specifically, the NUSW model, which accounts for both phase and power variations across array elements, demonstrates diminishing returns for large $N$, which is more accurate than the UPW and USW models. However, due to its failure to consider aperture loss and polarization loss, the CRs and SRs of the NUSW model still exhibits a slow but persistent increase with the number of array elements, which is not feasible in practical scenarios. Therefore, though the USW and NUSW models might find application in certain near-field scenarios, they are not applicable for scenarios with a large number of antennas. Further, it can be observed from Fig.~\ref{do_CRvsN} and Fig.~\ref{do_SRvsN} that both CRs and SRs for the scenario where the polarization mismatch is avoided also converge to finite limits as $N$ increases, which is consistent with our analysis. The S\&C performance achieved under this scenario is larger than performance in the case with polarization loss, establishing an ideal performance upper bound.  

By setting $r_{\mathrm{s}}=r$ and $r_{\mathrm{c}}=2r$, Fig.~\ref{do_r} illustrate S\&C performance as it relates to $r$. We can observe that, for a given number of antennas, the performance gap between different models diminishes as the distances from the CU and target to the BS increase. This is because the effect of the near-field is pronounced at short distances, where the effective antenna apertures and polarization mismatches among the elements vary significantly. Consequently, the rates of the TCMs, which ignore such impacts, are markedly overestimated when the CU and target are located near the BS. As the distance extends, the CU and target move toward the far field where the near-field effect is alleviated, though effective aperture loss and polarization loss remain, resulting in a narrow but constant gap between the accurate model and the TCMs at large distance.

Fig.~\ref{do_region} presents the downlink SR-CR regions achieved by the two systems: ISAC system (as defined in \eqref{Rate_Regio_ISAC}) and the baseline FDSAC system (as defined in \eqref{Rate_Regio_FDSAC}). On the graph, we can observe two marked points representing the S-C and C-C designs, respectively. The curve connecting these two points signifies the Pareto boundary of the downlink ISAC's rate region, which was obtained by solving the problem \eqref{Problem_CR_SR_Tradeoff} for values of $\sigma$ ranging from $1$ to $0$. It's essential to emphasize that the rate region attained by downlink FDSAC is entirely encompassed within the rate region of ISAC, thus validating \textbf{Theorem}~\ref{rate_region}. Furthermore, we also observe that the Pareto boundary achieved by the beamformer outlined in \textbf{Theorem}~\ref{pareto_theo} perfectly coincides with the boundary obtained from the beamformer presented in \textbf{Corollary}~\ref{pareto_cor}, which provides further support for the conclusion presented in \textbf{Remark}~\ref{pareto_rem}.

\begin{figure*}[!t]
    \centering
    \subfigbottomskip=5pt
	\subfigcapskip=0pt
\setlength{\abovecaptionskip}{0pt}
    \subfigure[CR vs. $N$ with $p_{\mathrm{c}}=85$ dB and $p_{\mathrm{s}}=60$ dB.]
    {
        \includegraphics[height=0.28\textwidth]{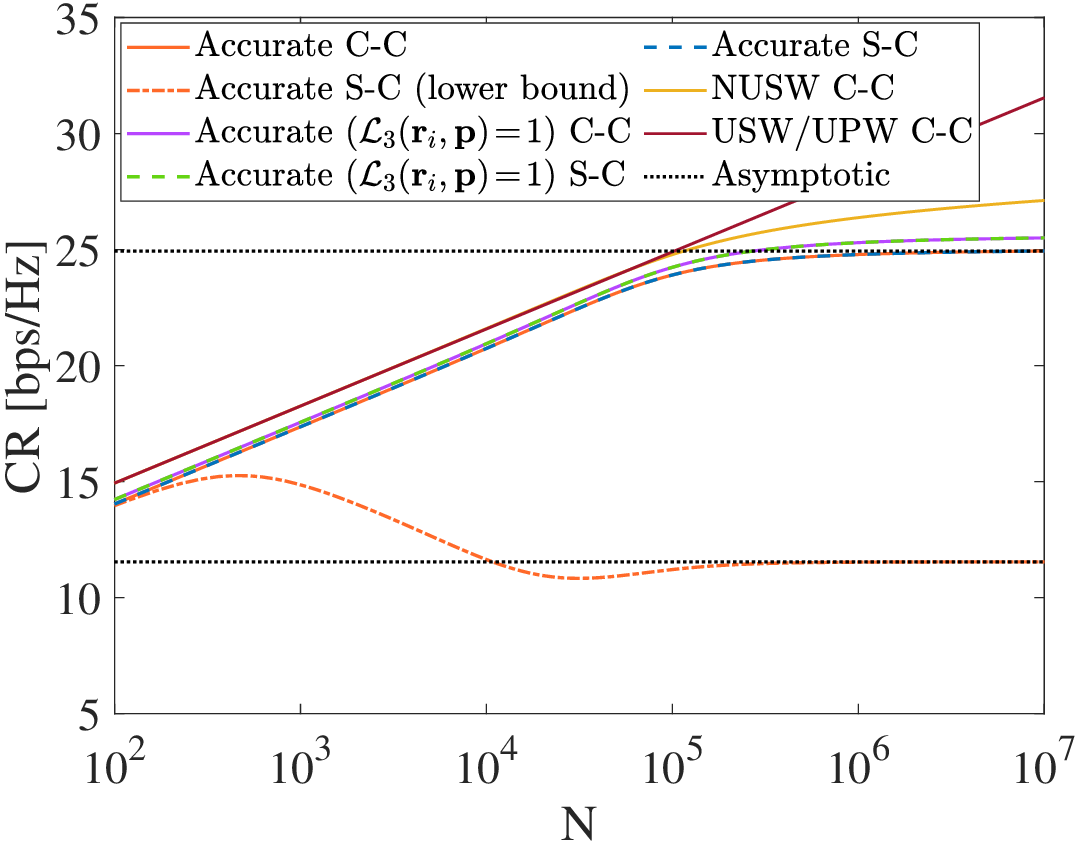}
	   \label{up_CRvsN}	
    }
    \quad
   \subfigure[SR vs. $N$ with $p_{\mathrm{s}}=85$ dB and $p_{\mathrm{c}}=60$ dB.]
    {
        \includegraphics[height=0.28\textwidth]{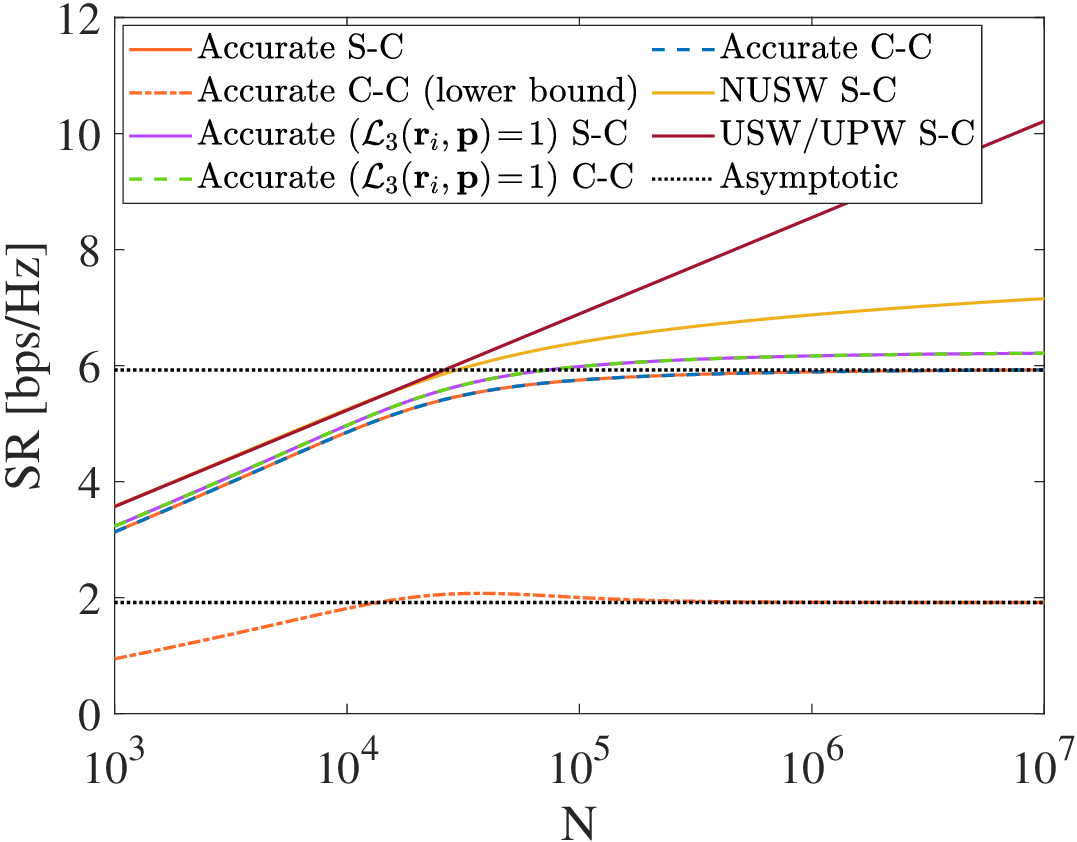}
	   \label{up_SRvsN}	
    }
\caption{Uplink performance versus number of array elements.}
    \label{up_N}
\end{figure*}

\subsection{Uplink}
For the uplink results, we first focus on Fig.~\ref{up_CRvsp} and Fig.~\ref{up_SRvsp}, which demonstrates the uplink CR and SR concerning the transmit SNR $p_{\mathrm{c}}$ and $p_{\mathrm{s}}$, respectively. As anticipated, the C-C ISAC exhibits the highest communication performance, while S-C ISAC achieves the highest SR. Remarkably, C-C ISAC and S-C ISAC exhibit the same high-SNR slopes, which outperform those achieved by FDSAC, as stated in \textbf{Remark}~\ref{up_FDSAC_compare}. Furthermore, both for CR and SR, a consistent performance gap exists between S-C ISAC and C-C ISAC in the high-SNR region. This observation aligns with the discussions presented in \textbf{Remark}~\ref{up_compare}.

In Fig.~\ref{up_CRvsN} and Fig.~\ref{up_SRvsN}, we present the uplink CR and SR as functions of the number of antennas $N$, respectively. As mentioned before, the gaps of the CRs and SRs between the uplink C-C and S-C designs under our accurate model are negligible in the near-field region, which is consistent with the results in these two graphs. Thus, the figures also display the derived lower bounds for the CR of the S-C design and the SR of the C-C designs. Notably, for small values of $N$, the rates achieved in all models exhibit a linear increase with $\log N$. This is because, when $N$ is small, the CU/target can be treated as in the far field, where all models are accurate. However, when $N$ is sufficiently large, the disparity in effective antenna apertures and polarization mismatches across the UPA becomes significant. In this case, the rates of the TCMs are overestimated due to the ignorance of the above impacts and will grow unboundedly with $N$, breaking the law of energy conservation. By contrast, as $N$ approaches infinity, with the accurate model, the CRs and SRs, along with their aforementioned lower bounds and upper bounds achieved through mitigating polarization mismatch, are capped at finite values, justifying the accuracy of the near-field channel model proposed in our work.

\begin{figure} [!t]
\centering
\includegraphics[height=0.305\textwidth]{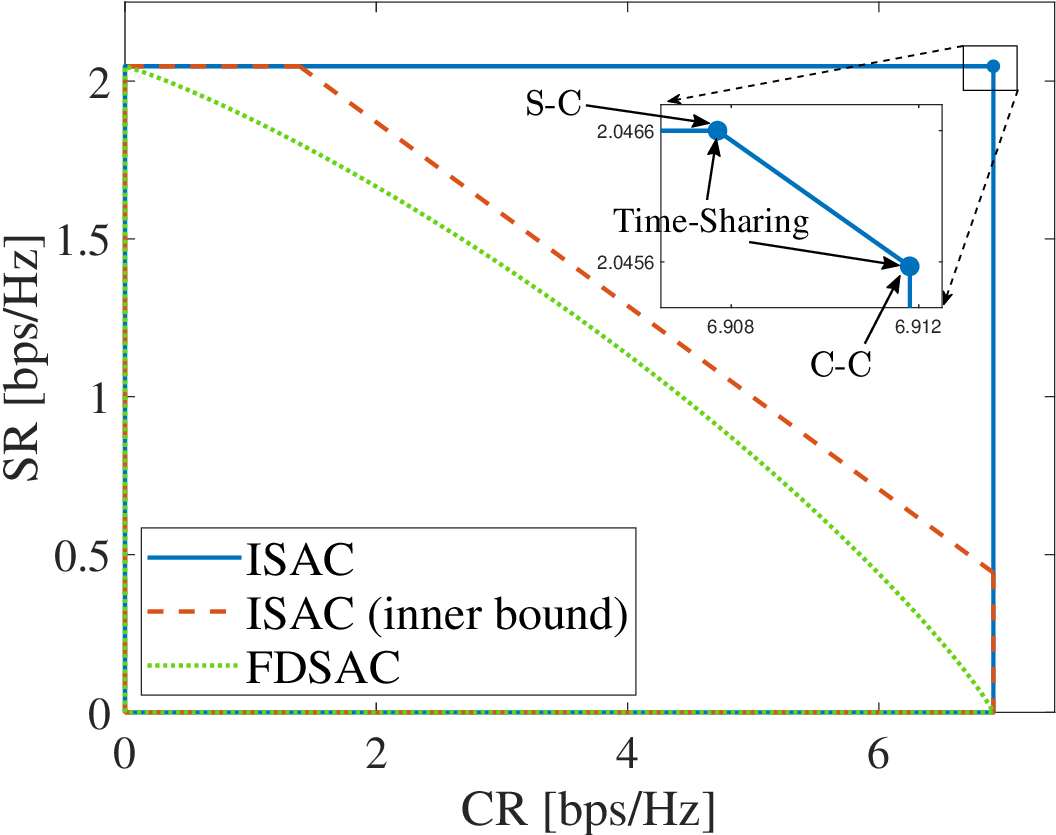}
 \caption{Uplink rate region with $p_{\mathrm{s}}=85$ dB and $p_{\mathrm{c}}=60$ dB.}
 \vspace{-5pt}
 \label{up_region}
\end{figure}

Fig.~\ref{up_region} illustrates the SR-CR regions attained by the uplink near-field FDSAC and ISAC systems. The two points on the plot correspond to the rates achieved by the S-C and C-C schemes, respectively, while the line segment connecting these points represents the rates attainable through a time-sharing strategy between the two schemes. The inner bound is achieved by the lower bounds of SR and CR as specified in \eqref{R_cs_bound} and \eqref{R_sc_bound}, respectively. A crucial observation from the plot is that the achievable rate region of the uplink FDSAC is wholly contained within that of the uplink ISAC and even its inner bound, illustrating the superiority of ISAC over FDSAC in the near-field region.

\section{Conclusion}\label{conclusion}
This paper {has} investigated the S\&C performance of a near-field ISAC system for both downlink and uplink scenarios. By incorporating the impacts of effective aperture and polarization loss, a more accurate channel model than the TCMs was employed in our investigation. The downlink ISAC was analyzed under three scenarios: S-C design, C-C design, and Pareto optimal design, while two different scenario based on the interference cancellation order were considered in uplink case. For each scenario, CRs, SRs and their high-SNR approximations were derived. To gain further insight into our near-field channel model, we also derived the asymptotic performance when the UPA of the BS has an infinite number of antennas. Furthermore, we characterized the attainable SR-CR rate regions of ISAC and the traditional FDSAC for both downlink and uplink scenarios. These results have demonstrated the superior S\&C performance of ISAC over FDSAC, and more importantly, have underscored the accuracy of the proposed near-field ISAC model.

\begin{appendices}
\section{Proof of Lemma~\ref{SR_lemma}}\label{Appendix:A}
\renewcommand{\theequation}{A.\arabic{equation}}
\setcounter{equation}{0} 
Vectorizing the sensing signal $\mathbf{Y}_{\mathrm{s}}$ in \eqref{echo}, we get
\begin{align}
\mathrm{vec}\left( \mathbf{Y}_{\mathrm{s}} \right) =\sqrt{p}\mathbf{h}_{\mathrm{s}}^{\mathsf{T}}\mathbf{w}\mathrm{vec}\left( \mathbf{h}_{\mathrm{s}}\mathbf{s}^{\mathsf{H}} \right) \beta +\mathrm{vec}\left( \mathbf{N}_{\mathrm{s}} \right) .    
\end{align} 
It is worth noting that the conditional MI between $\mathrm{vec}\left( \mathbf{Y}_{\mathrm{s}} \right)$ and $\beta$ is equivalent to the capacity of a MISO Gaussian channel with a Gaussian distributed input $\beta \sim \mathcal{C} \mathcal{N} \left( 0,\alpha _s \right) $: $\dot{\mathbf{y}}=\dot{\mathbf{h}}\beta +\dot{\mathbf{n}}$, where $\dot{\mathbf{h}}=\sqrt{p}\mathbf{h}_{\mathrm{s}}^{\mathsf{T}}\mathbf{w}\mathrm{vec}\left( \mathbf{h}_{\mathrm{s}}\mathbf{s}^{\mathsf{H}} \right) $ represents the channel vector, and $\dot{\mathbf{n}}\sim \mathcal{C} \mathcal{N} \left( \mathbf{0},\mathbf{I} \right) $. Therefore, the sensing MI can be calculated as $I\left( \mathbf{Y}_{\mathrm{s}};\beta |\mathbf{X} \right) =\log _2\det \left( \mathbf{I}+\dot{\mathbf{h}}\dot{\mathbf{h}}^{\mathsf{H}} \right)$. By applying Sylvester’s identity, we can further obtain
\begin{align} 
 I\left( \mathbf{Y}_{\mathrm{s}};\beta |\mathbf{X} \right) &=\log _2\left( 1+\dot{\mathbf{h}}^{\mathsf{H}}\dot{\mathbf{h}} \right)   \nonumber\\
 &=\log _2\left( 1+pL\alpha _{\mathrm{s}}\left\| \mathbf{h}_{\mathrm{s}} \right\| ^2\left| \mathbf{h}_{\mathrm{s}}^{\mathsf{T}}\mathbf{w} \right|^2 \right).\label{sensing_MI}    
\end{align}
Substituting \eqref{sensing_MI} into \eqref{SR_define}, we can obtain the final result.

\section{Proof of Theorem \ref{CC_CR_theorem}}\label{Appendix:B}
\renewcommand{\theequation}{B.\arabic{equation}}
\setcounter{equation}{0}
Based on \eqref{channel_model}, we can calculate $\left\| \mathbf{h}_{\mathrm{c}} \right\| ^2$ as 
\begin{align} \label{B1}
\left\| \mathbf{h}_{\mathrm{c}} \right\| ^2&=\frac{A\Psi _{\mathrm{c}}}{4\pi r_{\mathrm{c}}^{2}}\sum_{n_y=-\frac{N_y-1}{2}}^{\frac{N_y-1}{2}}{\sum_{n_z=-\frac{N_z-1}{2}}^{\frac{N_z-1}{2}}}\notag\\
 &\times \frac{\Psi_{\mathrm{c}} ^2+\left( \Omega_{\mathrm{c}} -n_z\epsilon _{\mathrm{c}} \right) ^2}{\left[ \left( n_y\epsilon _{\mathrm{c}}-\Phi _{\mathrm{c}} \right) ^2+\left( n_z\epsilon _{\mathrm{c}}-\Omega _{\mathrm{c}} \right) ^2+\Psi_{\mathrm{c}} ^2 \right] ^{\frac{5}{2}}}.   
\end{align}
We define the function $f\left( y,z \right) \triangleq \frac{\Psi ^2+\left( \Omega -z \right) ^2}{\left[ \left( y-\Phi _{\mathrm{c}} \right) ^2+\left( z-\Omega _{\mathrm{c}} \right) ^2+\Psi ^2 \right] ^{\frac{5}{2}}}$ in the rectangular area $\mathcal{H} =\left\{ \left( y,z \right) \mid -\frac{N_y\epsilon _{\mathrm{c}}}{2}\leqslant y\leqslant \frac{N_y\epsilon _{\mathrm{c}}}{2},-\frac{N_z\epsilon _{\mathrm{c}}}{2}\leqslant z\leqslant \frac{N_z\epsilon _{\mathrm{c}}}{2} \right\}  $ that is then partitioned into $N_yN_z$ sub-rectangles, each with equal area $\epsilon _{\mathrm{c}}^2$. Since $\epsilon _{\mathrm{c}}\ll 1$, we have $f\left( y,z \right) \approx f\left( n_y\epsilon _{\mathrm{c}},n_z\epsilon _{\mathrm{c}} \right) $ for $\forall \left( y,z \right) \in \left\{ \left( y,z \right) \mid \left( n_y-\frac{1}{2} \right) \epsilon _{\mathrm{c}}\leqslant y\leqslant \left( n_y+\frac{1}{2} \right) \epsilon _{\mathrm{c}},\left( n_z-\frac{1}{2} \right) \epsilon _{\mathrm{c}}\leqslant z\leqslant \right.$ $\left.\left( n_z+\frac{1}{2} \right) \epsilon _{\mathrm{c}} \right\}$. Based on the concept of integral, we have $\sum_{n_y=-\frac{N_y-1}{2}}^{\frac{N_y-1}{2}}{\sum_{n_z=-\frac{N_z-1}{2}}^{\frac{N_z-1}{2}}{f\left( n_y\epsilon _{\mathrm{c}},n_z\epsilon _{\mathrm{c}} \right) \epsilon _{\mathrm{c}}^2}}\approx \iint_{\mathcal{H}}{f\left( y,z \right) dydz}$. Therefore, \eqref{B1} can be rewritten as 
\begin{align}
\left\| \mathbf{h}_{\mathrm{c}} \right\| ^2&=\frac{\zeta \Psi _{\mathrm{c}}}{4\pi}\int_{-\frac{N_z\epsilon _{\mathrm{c}}}{2}}^{\frac{N_z\epsilon _{\mathrm{c}}}{2}}{\int_{-\frac{N_y\epsilon _{\mathrm{c}}}{2}}^{\frac{N_y\epsilon _{\mathrm{c}}}{2}}}\notag\\
&\times \frac{\Psi_{\mathrm{c}} ^2+\left( \Omega_{\mathrm{c}} -z \right) ^2}{\left[ \left( y-\Phi _{\mathrm{c}} \right) ^2+\left( z-\Omega _{\mathrm{c}} \right) ^2+\Psi_{\mathrm{c}} ^2 \right] ^{\frac{5}{2}}}dydz.   
\end{align}
We can calculate the inner integral with the aid of \cite[Eq. (2.263.3) \& (2.264.5)]{integral} and then the outer integral with the aid of \cite[Eq. (2.284.5)]{integral}, which yields
\begin{align}\label{b5}
\left\| \mathbf{h}_{\mathrm{c}} \right\| ^2=\frac{\zeta}{4\pi} \sum_{y\in \mathcal{Y}_{\mathrm{c}}}{\sum_{z\in \mathcal{Z}_{\mathrm{c}}}{\delta _{\mathrm{c}}\left( y,z \right)}} .
\end{align}
Substituting \eqref{b5} into \eqref{eq_Rcc}, we obtain the final results in \eqref{CC_CR}. 

For the high-SNR approximation, by applying the fact of $\lim_{x\rightarrow\infty}\log_2(1+x)\approx\log_2{x}$ to \eqref{CC_CR}, we can easily get \eqref{CC_CR_p}.

\section{Proof of Theorem \ref{rcc_polar_cor}}\label{Appendix:ex}
\renewcommand{\theequation}{C.\arabic{equation}}
\setcounter{equation}{0}
Following steps similar to those in Appendix~\ref{Appendix:B}, we can obtain
\begin{align}
\left\| \bar{\mathbf{h}}_{\mathrm{c}} \right\| ^2&=\frac{\zeta \Psi _{\mathrm{c}}}{4\pi}\int_{-\frac{N_z\epsilon _{\mathrm{c}}}{2}}^{\frac{N_z\epsilon _{\mathrm{c}}}{2}}{\int_{-\frac{N_y\epsilon _{\mathrm{c}}}{2}}^{\frac{N_y\epsilon _{\mathrm{c}}}{2}}}\notag\\
&\times (y^2+z^2-2\Phi_{\mathrm{c}}y-2\Omega_{\mathrm{c}}z+1)^{-\frac{3}{2}}dydz.   
\end{align}
We can calculate the inner integral with the aid of \cite[Eq. (2.264.5)]{integral} and then the outer integral with the aid of \cite[Eq. (2.284.5)]{integral}, which yields the results of \eqref{rcc_polar}. 
The asymptotic results for $N_y,N_z \rightarrow\infty$ can be derived based on $\lim_{y,z\rightarrow \infty} \arctan \left( \frac{yz}{\Psi _{\mathrm{c}}\sqrt{\Psi _{\mathrm{c}}^{2}+y^2+z^2}} \right) =\frac{\pi}{2}$.

\section{Proof of Theorem \ref{pareto_theo}}\label{Appendix:C}
\renewcommand{\theequation}{D.\arabic{equation}}
\setcounter{equation}{0}
The optimal solution of problem \eqref{Problem_CR_SR_Tradeoff} can be obtained from the Karush-Kuhn-Tucker (KKT) condition as follows:
\begin{numcases}{}
\nabla \left( -\mathcal{R} \right) +\lambda \nabla \left( \left\| \mathbf{w} \right\| ^2-1 \right) +\mu _1\nabla f_1+\mu _2\nabla f_2=\mathbf{0},\label{c1}\\
\mu _1f_1=0,\ \mu _2f_2=0,\ \mu _1\geqslant 0,\ \mu _2\geqslant 0,
\end{numcases}
where $f_1=2^{\left( 1-\sigma \right) \mathcal{R}}-1-p\left| \mathbf{h}_{\mathrm{c}}^{\mathsf{T}}\mathbf{w}\right|^2$, $f_2=2^{ \sigma L  \mathcal{R}}-1-p\xi^2\left| \mathbf{h}_{\mathrm{s}}^{\mathsf{T}}\mathbf{w} \right|^2$ and $\lambda$, $\mu_1$, $\mu_2$ are real Lagrangian multipliers. From \eqref{c1} and the constraint $\left\| \mathbf{w} \right\| ^2=1$, we obtain 
\begin{numcases}{}
\mu _1p\mathbf{h}_{\mathrm{c}}\mathbf{h}_{\mathrm{c}}^{\mathsf{H}}\mathbf{w}+\mu _2p\xi^2\mathbf{h}_{\mathrm{s}}\mathbf{h}_{\mathrm{s}}^{\mathsf{H}}\mathbf{w}=\lambda \mathbf{w},\label{c3}\\
\mu _1p\left| \mathbf{h}_{\mathrm{c}}^{\mathsf{T}}\mathbf{w} \right|^2+\mu _2p\xi^2\left| \mathbf{h}_{\mathrm{s}}^{\mathsf{T}}\mathbf{w} \right|^2=\lambda \mathbf{w}^{\mathsf{H}}\mathbf{w}=\lambda ,\label{c4}\\
\mu _1 2^{\left( 1-\sigma \right)\mathcal{R}}\left( 1-\sigma \right) \ln 2+\mu _2 2^{\sigma L\mathcal{R}}\sigma L \ln 2=1.\label{c5}
\end{numcases}
It follows from \eqref{c5} that $\mu_1$ and $\mu_2$ cannot be $0$ at the same time. Subsequently, we discuss three cases as follows. 

\subsubsection{$\mu _1>0$ \& $\mu _2=0$}
In this case, we have
\begin{numcases}{}
\mu _1p\mathbf{h}_{\mathrm{c}}\mathbf{h}_{\mathrm{c}}^{\mathsf{H}}\mathbf{w}=\lambda \mathbf{w},\label{c6}\\
\mu _1p\left| \mathbf{h}_{\mathrm{c}}^{\mathsf{T}}\mathbf{w} \right|^2=\lambda ,\label{c7}\\
\mu _1 2^{\left( 1-\sigma \right)\mathcal{R}}\left( 1-\sigma \right)\ln 2=1,\label{c8}\\
f_2=0\Rightarrow p\left| \mathbf{h}_{\mathrm{c}}^{\mathsf{T}}\mathbf{w} \right|^2= 2^{\left( 1-\sigma \right) \mathcal{R}}-1.  \label{c9}  
\end{numcases}
According to \eqref{c9}, to maximize $\mathcal{R}$, we can obtain the optimal beamforming vector as $\mathbf{w}_\sigma^{\star}=\mathbf{w}_{\mathrm{c}}$, which is followed by \begin{align}
\mathcal{R} ^{\star}=\frac{1}{1-\sigma}\log \left( 1+p\left\| \mathbf{h}_{\mathrm{c}} \right\| ^2 \right) =\frac{\mathcal{R} _{\mathrm{d},\mathrm{c}}^{\mathrm{c}}}{1-\sigma}.    
\end{align}
Under this circumstance, we have $\mathcal{R} _{\mathrm{d},\mathrm{s}}=\mathcal{R} _{\mathrm{d},\mathrm{s}}^{\mathrm{c}} \geqslant \sigma \mathcal{R}^{\star}$, which yields $\sigma \in \left[ 0,\frac{\mathcal{R} _{\mathrm{d},\mathrm{s}}^{\mathrm{c}}}{\mathcal{R} _{\mathrm{d},\mathrm{c}}^{\mathrm{c}}+\mathcal{R} _{\mathrm{d},\mathrm{s}}^{\mathrm{c}}} \right] $.

\subsubsection{$\mu _1=0$ \& $\mu _2>0$}
Following the similar steps in the first case, we can obtain $\mathbf{w}_\sigma^{\star}=\mathbf{w}_{\mathrm{s}}$ and $\mathcal{R} ^{\star}=\frac{\mathcal{R} _{\mathrm{d},\mathrm{s}}^{\mathrm{s}}}{\sigma}$, for $\sigma \in \left[ \frac{\mathcal{R} _{\mathrm{d},\mathrm{s}}^{\mathrm{s}}}{\mathcal{R} _{\mathrm{d},\mathrm{c}}^{\mathrm{s}}+\mathcal{R} _{\mathrm{d},\mathrm{s}}^{\mathrm{s}}},1 \right] $.

\subsubsection{$\mu _1>0$ \& $\mu _2>0$}
In this case, we have 
\begin{equation}\label{c10}
\begin{cases}
f_1=0\Rightarrow p\left| \mathbf{h}_{\mathrm{c}}^{\mathsf{T}}\mathbf{w} \right|^2= 2^{\left( 1-\sigma \right) \mathcal{R}}-1,\\
 f_2=0\Rightarrow p\xi^2\left| \mathbf{h}_{\mathrm{s}}^{\mathsf{T}}\mathbf{w} \right|^2= 2^{\sigma L \mathcal{R}}-1.
\end{cases}
\end{equation}
From \eqref{c3}, we can write $\mathbf{w}$ as follows:
\begin{align}\label{c11}
\mathbf{w}&=\frac{\mu _1p\mathbf{h}_{\mathrm{c}}^{\mathsf{H}}\mathbf{w}}{\lambda}\mathbf{h}_{\mathrm{c}}+\frac{\mu _2p\xi^2\mathbf{h}_{\mathrm{s}}^{\mathsf{H}}\mathbf{w}}{\lambda}\mathbf{h}_{\mathrm{s}}\triangleq a\mathbf{h}_{\mathrm{c}}+b\mathbf{h}_{\mathrm{s}}.    
\end{align}
Substituting \eqref{c11} into \eqref{c3} gives
\begin{align}\label{c12}
\mu _1p\Big( \left\| \mathbf{h}_{\mathrm{c}} \right\| ^2+\frac{b}{a} \mathbf{h}_{\mathrm{c}}^{\mathsf{H}}\mathbf{h}_{\mathrm{s}} \Big) =\mu _2p\xi^2 \left( \left\| \mathbf{h}_{\mathrm{s}} \right\| ^2+\frac{a}{b}\mathbf{h}_{\mathrm{s}}^{\mathsf{H}}\mathbf{h}_{\mathrm{c}} \right) =\lambda ,
\end{align}
where $\frac{a}{b}=\frac{\mu _1\chi^{-1}}{\mu _2\xi^2{\rm{e}}^{-{\rm{j}}\angle(\mathbf{h}_{\mathrm{c}}^{\mathsf{H}}\mathbf{h}_{\mathrm{s}})}}$ is obtained according to \eqref{c10}. By combining \eqref{c12} and \eqref{c5}, we can derive the expressions of $\mu_1$, $\mu_2$ and $\lambda$. Substituting these expressions into \eqref{c4}, we can obtain the equation \eqref{equation_R} for $\mathcal{R}$. Since the left-hand side of \eqref{equation_R} is a monotonic function with respect to $\mathcal{R}$, ranging from $0$ to $\infty $, and the right-hand side is non-negative, a solution of the equation can be definitely found, which is followed by the results of $\mathbf{w}_\sigma^\star$.

\section{Proof of Corollary \ref{pareto_cor}}\label{Appendix:D}
\renewcommand{\theequation}{E.\arabic{equation}}
\setcounter{equation}{0}
The attainable SR-CR regions achieved by $\mathbf{w}_{\sigma}^{\star}$ and $\mathbf{w}_{\tau}$ are given by
\begin{align}
&\!\!\mathcal{C}_{\sigma}\!=\!\left\{\left({\mathcal{R}}_{\mathrm{s}},{\mathcal{R}}_{\mathrm{c}}\right)|{\mathcal{R}}_{\mathrm{s}}\!\in\!\left[0,\mathcal{R}_{\mathrm{s}}^{\sigma}\right],
{\mathcal{R}}_{\mathrm{c}}\!\in\!\left[0,\mathcal{R}_{\mathrm{c}}^{\sigma}\right],\sigma\!\in\!\left[0,1\right]\right\},\\
&\!\!\mathcal{C}_{\tau}\!=\!\left\{\left({\mathcal{R}}_{\mathrm{s}},{\mathcal{R}}_{\mathrm{c}}\right)|{\mathcal{R}}_{\mathrm{s}}\!\in\!\left[0,\mathcal{R}_{\mathrm{s}}^{\tau}\right],
{\mathcal{R}}_{\mathrm{c}}\!\in\!\left[0,\mathcal{R}_{\mathrm{c}}^{\tau}\right],\tau\!\in\!\left[0,1\right]\right\},
\end{align}
where $\left( \mathcal{R} _{\mathrm{s}}^{\sigma},\mathcal{R} _{\mathrm{c}}^{\sigma} \right) $ and $\left( \mathcal{R} _{\mathrm{s}}^{\tau},\mathcal{R} _{\mathrm{c}}^{\tau} \right) $ denote the rate pairs achieved by $\mathbf{w}_{\sigma}^{\star}$ and $\mathbf{w}_{\tau}$, respectively. Since $\mathcal{C}_{\sigma}$ encompasses all achievable rate pairs, we have $\mathcal{C} _{\tau}\subseteq \mathcal{C} _{\sigma}$. Furthermore, because $\mathbf{w}_{\sigma}^{\star}$ is the linear combination of $\mathbf{h}_{\mathrm{c}}$ and $\mathbf{h}_{\mathrm{s}} {\rm{e}}^{-{\rm{j}}\angle \psi}$ with non-negative real coefficients, and $\mathbf{w}_{\tau}$ can represent any arbitrary linear combination of $\mathbf{h}_{\mathrm{c}}$ and $\mathbf{h}_{\mathrm{s}} {\rm{e}}^{-{\rm{j}}\angle \psi}$ with non-negative real coefficients, we have $\mathcal{C} _{\sigma}\subseteq \mathcal{C} _{\tau}$. Consequently, we obtain $\mathcal{C} _{\sigma}=\mathcal{C} _{\tau}$, leading to the results presented in Corollary~\ref{pareto_cor}.

\section{Proof of Theorem \ref{rate_region}}\label{Proof_rate_region}
\renewcommand{\theequation}{F.\arabic{equation}}
\setcounter{equation}{0}
Firstly, We define an auxiliary region as follow:
\begin{align}
\!\!\mathcal{C} _a\!\!=\!\!\left\{ \!\left( \mathcal{R} _{\mathrm{s}},\mathcal{R} _{\mathrm{c}} \right)\! |\mathcal{R} _{\mathrm{s}}\!\in \!\!\left[ 0,\mathcal{R} _{a,\mathrm{s}}^{\varsigma } \right] \!,\mathcal{R} _{\mathrm{c}}\!\in \!\!\left[ 0,\mathcal{R} _{a,\mathrm{c}}^{\varsigma } \right] \!,\varsigma  \!\in \!\!\left[ 0,1 \right] \right\} \!,
\end{align}
where $\mathcal{R} _{a,\mathrm{s}}^{\varsigma }$ and $\mathcal{R} _{a,\mathrm{c}}^{\varsigma }$ are, respectively, defined as $\mathcal{R} _{a,\mathrm{s}}^{\varsigma }=\frac{1}{L}\log _2\left( 1+\varsigma  pL\alpha _{\mathrm{s}}\left\| \mathbf{h}_{\mathrm{s}} \right\| ^4 \right)$ and $\mathcal{R} _{a,\mathrm{c}}^{\varsigma }=\log _2\left( 1+\left( 1-\varsigma  \right) p\left\| \mathbf{h}_{\mathrm{c}}\right\| ^2 \right)$. As $\mathcal{C} _a$ is achieved by allocating power separately for communication and sensing while utilizing the entire bandwidth for each purpose, we have $\mathcal{C} _{\mathrm{d},\mathrm{f}}\subseteq \mathcal{C} _{a}$. It noteworthy that $\mathcal{R} _{a,\mathrm{s}}^{\varsigma }$ monotonically increases with $\varsigma $, while $\mathcal{R} _{a,\mathrm{c}}^{\varsigma }$ monotonically decreases with $\varsigma $. When $\varsigma =0$, we have $\mathcal{R} _{a,\mathrm{c}}^{0}= \mathcal{R} _{\mathrm{d},\mathrm{c}}^{\mathrm{c}} = \mathcal{R} _{\mathrm{d},\mathrm{c}}^{0}$ and $\mathcal{R} _{a,\mathrm{s}}^{0}=0< \mathcal{R} _{\mathrm{d},\mathrm{s}}^{\mathrm{c}} = \mathcal{R} _{\mathrm{d},\mathrm{s}}^{0}$. When $\varsigma =1$, we have $\mathcal{R} _{a,\mathrm{s}}^{1}= \mathcal{R} _{\mathrm{d},\mathrm{s}}^{\mathrm{s}} = \mathcal{R} _{\mathrm{d},\mathrm{s}}^{1}$ and $\mathcal{R} _{a,\mathrm{c}}^{1}=0< \mathcal{R} _{\mathrm{d},\mathrm{c}}^{\mathrm{s}} = \mathcal{R} _{\mathrm{d},\mathrm{c}}^{1}$. Therefore, it is easily shown that $\mathcal{C} _a\subseteq \mathcal{C} _{\mathrm{d},\mathrm{i}}$. Consequently, we obtain $\mathcal{C} _{\mathrm{d},\mathrm{f}}\subseteq \mathcal{C} _{a}\subseteq \mathcal{C} _{\mathrm{d},\mathrm{i}}$.

\section{Proof of Theorem~\ref{up_CC_SR_the}}\label{Appendix:F}
\renewcommand{\theequation}{G.\arabic{equation}}
\setcounter{equation}{0}
Substituting \eqref{dual_function_signal_matrix}
and \eqref{G_model} into \eqref{BS_receive}, the received signal of the BS can be written as $\mathbf{Y}=\beta \mathbf{h}_{\mathrm{s}}\mathbf{h}_{\mathrm{s}}^{\mathsf{T}}\mathbf{ws}_{\mathrm{s}}^{\mathsf{H}}+\mathbf{Z}_{\mathrm{c}}$. Vectorizing the signal, we have
\begin{align}\label{f1}
\mathrm{vec}\left( \mathbf{Y} \right) =\sqrt{p}\mathbf{h}_{\mathrm{s}}^{\mathsf{T}}\mathbf{w}\mathrm{vec}\left( \mathbf{h}_{\mathrm{s}}\mathbf{s}_{\mathrm{s}}^{\mathsf{H}} \right) \beta +\mathbf{z}_{\mathrm{c}},    
\end{align}
where $\mathbf{z}_{\mathrm{c}}=\mathrm{vec}\left( \mathbf{Z}_{\mathrm{c}} \right) $. By regarding \eqref{f1} as a MISO channel model with a zero-mean Gaussian noise $\mathbf{z}_{\mathrm{c}}$, the maximum SR with $\mathbf{w}=\mathbf{w}_{\mathrm{s}}$ can be calculated as follows:
\begin{align}\label{f2}
\mathcal{R} _{\mathrm{c},\mathrm{s}}^{\mathrm{c}}\!=\!\frac{1}{L}\log _2\!\left[ 1\!+\!\alpha_{\mathrm{s}}p_{\mathrm{s}}\left\| \mathbf{h}_{\mathrm{s}} \right\|^2\mathrm{vec}^{\mathsf{H}}\!\left( \mathbf{h}_{\mathrm{s}}\mathbf{s}_{\mathrm{s}}^{\mathsf{H}} \right) \mathbf{R}_{\mathrm{c}}^{-1}\mathrm{vec}\!\left( \mathbf{h}_{\mathrm{s}}\mathbf{s}_{\mathrm{s}}^{\mathsf{H}} \right) \right] ,    
\end{align}
where $\mathbf{R}_{\mathrm{c}}=\mathbb{E} \left\{ \mathbf{z}_{\mathrm{c}}\mathbf{z}_{\mathrm{c}}^{\mathsf{H}} \right\} $. Since we have $\mathbb{E} \left\{ \mathbf{s}_{\mathrm{c}}\mathbf{s}_{\mathrm{c}}^{\mathsf{H}} \right\} =\mathbf{I}_L$ and $\mathbf{N}_{\mathrm{u}}\sim \mathcal{C} \mathcal{N} \left( \mathbf{0},\mathbf{I} \right) $, $\mathbf{R}_{\mathrm{c}}$ is a block diagonal matrix composed of $L$ blocks of $\mathbf{A}\!=\!p_{\mathrm{c}}\mathbf{h}_{\mathrm{c}}\mathbf{h}_{\mathrm{c}}^{\mathsf{H}}\!+\!\mathbf{I}_N$. Thus, \eqref{f2} can be rewritten as
\begin{align}\label{f3}
\mathcal{R} _{\mathrm{c},\mathrm{s}}^{\mathrm{c}}=\frac{1}{L}\log _2\left( 1+p_{\mathrm{s}}L\alpha _{\mathrm{s}}\left\| \mathbf{h}_{\mathrm{s}} \right\| ^2\mathbf{h}_{\mathrm{s}}^{\mathsf{H}}\mathbf{A}^{-1}\mathbf{h}_{\mathrm{s}} \right).     
\end{align}
By applying the Woodbury matrix identity, we can obtain
\begin{align}\label{f4}
\mathbf{A}^{-1}\!=\!\left( p_{\mathrm{c}}\mathbf{h}_{\mathrm{c}}\mathbf{h}_{\mathrm{c}}^{\mathsf{H}}\!+\mathbf{I}_N \right) ^{-1}\!=\mathbf{I}_N-p_u\mathbf{h}_{\mathrm{c}}\!\left( 1\!+\!p_{\mathrm{c}}\left\| \mathbf{h}_{\mathrm{c}} \right\| ^2 \right) ^{-1}\!\mathbf{h}_{\mathrm{c}}^{\mathsf{H}}.
\end{align}
Substituting \eqref{f4} into \eqref{f3} yields the results in Theorem~\ref{up_CC_SR_the}.

\section{Proof of Theorem~\ref{up_SC_CR_the}}\label{Appendix:G}
\renewcommand{\theequation}{H.\arabic{equation}}
\setcounter{equation}{0}
Since the CR remains constant across all time slots, without loss of generality, we focus on the $l$th time slot with $l\in \left\{ 1,\ldots,L\right\}$. Under the S-C design, to reach the optimal SR, we have $\mathbf{w}=\mathbf{w}_{\mathrm{s}}$, and thus the received signal of BS at $l$th time slot is given by
\begin{align}
\mathbf{y}_l=\sqrt{p_{\mathrm{c}}}\mathbf{h}_{\mathrm{c}}\mathrm{s}_{\mathrm{c},1}+\underset{\mathbf{z}_{\mathrm{s},l}}{\underbrace{\sqrt{p}\beta \left\| \mathbf{h}_{\mathrm{s}} \right\| \mathbf{h}_{\mathrm{s}}\mathrm{s}_{\mathrm{s},l}+\mathbf{n}_{\mathrm{u},l}}},    
\end{align}
where $\mathbf{z}_{\mathrm{s},l}$ is a zero-mean Gaussian noise. In this case, the maximum achievable CR is calculated as follows:
\begin{align}\label{g2}
\mathcal{R} _{\mathrm{c},\mathrm{c}}^{\mathrm{s}}=\log _2\left( 1+p_{\mathrm{c}}\mathbf{h}_{\mathrm{c}}^{\mathsf{H}}\mathbf{R}_{\mathrm{s}}^{-1}\mathbf{h}_{\mathrm{c}} \right),
\end{align}
where $\mathbf{R}_{\mathrm{s}}=\mathbb{E} \left\{ \mathbf{z}_{\mathrm{s}}\mathbf{z}_{\mathrm{s}}^{\mathsf{H}} \right\} =p_{\mathrm{s}}\alpha _{\mathrm{s}}\left\| \mathbf{h}_{\mathrm{s}} \right\| ^2\mathbf{h}_{\mathrm{s}}\mathbf{h}_{\mathrm{s}}^{\mathsf{H}}+\mathbf{I}_N$. By applying the Woodbury matrix identity, $\mathbf{R}_{\mathrm{s}}^{-1}$ can be written as
\begin{align}\label{g3}
\mathbf{R}_{\mathrm{s}}^{-1}=\mathbf{I}_N-p_{\mathrm{s}}\alpha _{\mathrm{s}}\left\| \mathbf{h}_{\mathrm{s}} \right\| ^2\mathbf{h}_{\mathrm{s}}\left( 1+p_{\mathrm{s}}\alpha _{\mathrm{s}}\left\| \mathbf{h}_{\mathrm{s}} \right\| ^4 \right) ^{-1}\!\mathbf{h}_{\mathrm{s}}^{\mathsf{H}}.
\end{align}
Substituting \eqref{g3} into \eqref{g2} yields the results in Theorem~\ref{up_SC_CR_the}.

\end{appendices}

\bibliographystyle{IEEEtran}
\bibliography{IEEEabrv}

\begin{thebibliography}{10}
\providecommand{\url}[1]{#1}
\csname url@samestyle\endcsname
\providecommand{\newblock}{\relax}
\providecommand{\bibinfo}[2]{#2}
\providecommand{\BIBentrySTDinterwordspacing}{\spaceskip=0pt\relax}
\providecommand{\BIBentryALTinterwordstretchfactor}{4}
\providecommand{\BIBentryALTinterwordspacing}{\spaceskip=\fontdimen2\font plus
\BIBentryALTinterwordstretchfactor\fontdimen3\font minus \fontdimen4\font\relax}
\providecommand{\BIBforeignlanguage}[2]{{%
\expandafter\ifx\csname l@#1\endcsname\relax
\typeout{** WARNING: IEEEtran.bst: No hyphenation pattern has been}%
\typeout{** loaded for the language `#1'. Using the pattern for}%
\typeout{** the default language instead.}%
\else
\language=\csname l@#1\endcsname
\fi
#2}}
\providecommand{\BIBdecl}{\relax}
\BIBdecl

\bibitem{ICC_2024}
B.~Zhao, C.~Ouyang, X.~Zhang, and Y.~Liu, ``Performance analysis of near-field {ISAC} based on an accurate channel model,'' in \emph{Proc. IEEE Int. Conf. Commun. (ICC)}, 2024, pp. 1--6.

\bibitem{Zhang2021_JSTSP}
J.~A. Zhang \emph{et~al.}, ``An overview of signal processing techniques for joint communication and radar sensing,'' \emph{IEEE J. Sel. Topics Signal Process.}, vol.~15, no.~6, pp. 1295--1315, Nov. 2021.

\bibitem{ISACoverview_3}
Z.~Wei \emph{et~al.}, ``Integrated sensing and communication signals toward {5G-A} and {6G}: A survey,'' \emph{IEEE Internet Things J.}, vol.~10, no.~13, pp. 11\,068--11\,092, Jul. 2023.

\bibitem{LiuAn}
A.~Liu \emph{et~al.}, ``A survey on fundamental limits of integrated sensing and communication,'' \emph{IEEE Commun. Surveys Tuts.}, vol.~24, no.~2, pp. 994--1034, Feb. 2022.

\bibitem{star_1}
Y.~Liu \emph{et~al.}, ``{STAR}: Simultaneous transmission and reflection for 360° coverage by intelligent surfaces,'' \emph{{IEEE} Wireless Commun.}, vol.~28, no.~6, pp. 102--109, Dec. 2021.

\bibitem{star_2}
J.~Xu \emph{et~al.}, ``{STAR}-{RIS}s: Simultaneous transmitting and reflecting reconfigurable intelligent surfaces,'' \emph{{IEEE} Commun. Lett.}, vol.~25, no.~9, pp. 3134--3138, Sep. 2021.

\bibitem{starISAC}
Z.~Liu \emph{et~al.}, ``Toward {STAR}-{RIS}-empowered integrated sensing and communications: Joint active and passive beamforming design,'' \emph{{IEEE} Trans. Veh. Technol.}, pp. 1--15, early access, Jul. 11 2023.

\bibitem{Tang2019_TSP}
B.~Tang and J.~Li, ``Spectrally constrained {MIMO} radar waveform design based on mutual information,'' \emph{IEEE Trans. Signal Process.}, vol.~67, no.~3, pp. 821--834, Feb. 2019.

\bibitem{NFC}
Y.~Liu \emph{et~al.}, ``Near-field communications: A tutorial review,'' \emph{{IEEE} Open J.Commun. Soc.}, Early Access, Aug. 2023.

\bibitem{Xidong_magazine}
X.~Mu \emph{et~al.}, ``Reconfigurable intelligent surface-aided near-field communications for {6G}: Opportunities and challenges,'' \emph{{IEEE} Veh. Technol. Mag.}, pp. 2--11, early access, Jan. 04 2024.

\bibitem{rayleighdis}
J.~D. Kraus and R.~J. Marhefka, \emph{Antennas for All Applications}.\hskip 1em plus 0.5em minus 0.4em\relax New York, NY, USA: McGraw-Hill, 2002.

\bibitem{ISAC_performance1}
A.~R. Chiriyath, B.~Paul, G.~M. Jacyna, and D.~W. Bliss, ``Inner bounds on performance of radar and communications co-existence,'' \emph{{IEEE} Trans. Signal Process.}, vol.~64, no.~2, pp. 464--474, Jan. 2016.

\bibitem{ISAC_performance3}
M.~Liu \emph{et~al.}, ``Performance analysis and power allocation for cooperative {ISAC} networks,'' \emph{{IEEE} Internet Things J.}, vol.~10, no.~7, pp. 6336--6351, Apr. 2023.

\bibitem{Ouyang2022_WCL}
C.~Ouyang \emph{et~al.}, ``Performance of downlink and uplink integrated sensing and communications ({ISAC}) systems,'' \emph{IEEE Wireless Commun. Lett.}, vol.~11, no.~9, pp. 1850--1854, Sep. 2022.

\bibitem{boqun_NOMAISAC}
B.~Zhao \emph{et~al.}, ``Downlink and uplink {NOMA}-{ISAC} with signal alignment,'' arXiv preprint arXiv:2308.16352, 2023.

\bibitem{NFISAC_overview2}
J.~Cong \emph{et~al.}, ``Near-field integrated sensing and communication: Opportunities and challenges,'' arXiv preprint arXiv:2310.01342, 2023.

\bibitem{NFISAC_1}
Z.~Wang \emph{et~al.}, ``Near-field integrated sensing and communications,'' \emph{IEEE Commun. Lett.}, vol.~27, no.~8, pp. 2048--2052, Aug. 2023.

\bibitem{NFISAC_2}
K.~Qu, S.~Guo, and N.~Saeed, ``Near-field integrated sensing and communication: Performance analysis and beamforming design,'' arXiv preprint arXiv:2308.06455, 2023.

\bibitem{NFISAC_3}
H.~Luo \emph{et~al.}, ``Beam squint assisted user localization in near-field integrated sensing and communications systems,'' \emph{{IEEE} Trans. Wireless Commun.}, pp. 1--1, Early Access, Oct. 3 2023.

\bibitem{NFISAC_4}
A.~M. Elbir \emph{et~al.}, ``Near-field hybrid beamforming for terahertz-band integrated sensing and communications,'' arXiv preprint arXiv:2309.13984, 2023.

\bibitem{usw1}
D.~Starer \emph{et~al.}, ``Passive localization of near-field sources by path following,'' \emph{IEEE Trans. Signal Process.}, vol.~42, no.~3, pp. 677--680, Mar. 1994.

\bibitem{haiyang_TWC}
H.~Zhang \emph{et~al.}, ``Beam focusing for near-field multiuser {MIMO} communications,'' \emph{{IEEE} Trans. Wireless Commun.}, vol.~21, no.~9, pp. 7476--7490, Sep. 2022.

\bibitem{zengyong_twc}
H.~Lu and Y.~Zeng, ``Communicating with extremely large-scale array/surface: Unified modeling and performance analysis,'' \emph{IEEE Trans. Wireless Commun.}, vol.~21, no.~6, pp. 4039--4053, Jun. 2022.

\bibitem{polarization_1}
D.~Dardari, ``Communicating with large intelligent surfaces: Fundamental limits and models,'' \emph{IEEE J. Sel. Areas Commun.}, vol.~38, no.~11, pp. 2526--2537, Nov. 2020.

\bibitem{polarization_2}
E.~Björnson and L.~Sanguinetti, ``Power scaling laws and near-field behaviors of massive {MIMO} and intelligent reflecting surfaces,'' \emph{IEEE Open J. Commun. Soc.}, vol.~1, pp. 1306--1324, 2020.

\bibitem{uplink_SIC}
C.~Ouyang \emph{et~al.}, ``Revealing the impact of {SIC} in {NOMA}-{ISAC},'' \emph{{IEEE} Wireless Commun. Lett.}, vol.~12, no.~10, pp. 1707--1711, Oct. 2023.

\bibitem{polar_match}
S.~Chowdhury, J.~Frolik, and A.~Benslimane, ``Polarization matching for networks utilizing tripolar antenna systems,'' in \emph{Proc. IEEE Global Commun. Conf. (GLOBECOM)}, 2018, pp. 206--212.

\bibitem{multi_polar}
A.~Sousa~de Sena, D.~Benevides~da Costa, Z.~Ding, and P.~H.~J. Nardelli, ``Massive {MIMO}–{NOMA} networks with multi-polarized antennas,'' \emph{{IEEE} Trans. Wireless Commun.}, vol.~18, no.~12, pp. 5630--5642, Dec. 2019.

\bibitem{adapt_polar}
T.~Pratt, B.~Walkenhorst, and S.~Nguyen, ``Adaptive polarization transmission of {OFDM} signals in channels with polarization mode dispersion and polarization-dependent loss,'' \emph{{IEEE} Trans. Wireless Commun.}, vol.~8, no.~7, pp. 3354--3359, Jul. 2009.

\bibitem{pareto}
R.~Zhang and S.~Cui, ``Cooperative interference management with {MISO} beamforming,'' \emph{IEEE Trans. Signal Process.}, vol.~58, no.~10, pp. 5450--5458, Oct. 2010.

\bibitem{GaussianNoise}
B.~Hassibi \emph{et~al.}, ``How much training is needed in multiple-antenna wireless links?'' \emph{{IEEE} Trans. Inf. Theory}, vol.~49, no.~4, pp. 951--963, Apr 2003.

\bibitem{channel_correlation}
Z.~Wu and L.~Dai, ``Multiple access for near-field communications: {SDMA} or {LDMA}?'' \emph{IEEE J. Sel. Areas Commun.}, vol.~41, no.~6, pp. 1918--1935, Jun 2023.

\bibitem{mimo}
R.~W. Heath~Jr and A.~Lozano, \emph{Foundations of {MIMO} Communication}.\hskip 1em plus 0.5em minus 0.4em\relax Cambridge, U.K.: Cambridge Univ. Press, 2018.

\bibitem{integral}
I.~S. Gradshteyn and I.~M. Ryzhik, \emph{Table of Integrals, Series and Products}, 7th~ed.\hskip 1em plus 0.5em minus 0.4em\relax New York, NY, USA: Academic Press, 2007.

\end{thebibliography}

\end{document}